\documentclass{article}

\usepackage{geometry}
\usepackage{booktabs}   
\usepackage{listings}
\usepackage{comment}
\usepackage{graphicx}
\usepackage{balance}  
\usepackage{amsmath,amssymb,graphicx,multirow}
\usepackage[ruled,lined,noend,linesnumbered]{algorithm2e}
\usepackage{enumerate}
\usepackage{verbatim}
\usepackage{bm}
\usepackage{tabularx,times}
\usepackage{microtype}
\usepackage{color}
\usepackage{bbm}
\usepackage{arydshln}
\usepackage{url}
\usepackage[numbers]{natbib}
\newcolumntype{Y}{>{\centering\arraybackslash}X}

\let \originalleft \left
\let\originalright\right
\renewcommand{\left}{\mathopen{}\mathclose\bgroup\originalleft}
\renewcommand{\right}{\aftergroup\egroup\originalright}

\newcommand{\ctext}[1]{\texttt{\small #1}}

\newcommand{\ourname}{PAM}

\newcommand{\Tl}{T_L}
\newcommand{\Tr}{T_R}

\newcommand{\node}{\mbox{\tt Node}}
\newcommand{\leaf}{\mbox{\tt Leaf}}
\newcommand{\func}[1]{{\sc #1}}
\newcommand{\augtype}{{{A}}}
\newcommand{\augvalueone}[1]{{\mathcal{A}(#1)}}
\newcommand{\augvalue}[3]{{\mathcal{A}_{#1,#2}(#3)}}
\newcommand{\join}{\func{Join}}
\newcommand{\find}{\func{Find}}
\renewcommand{\split}{\func{Split}}
\newcommand{\augFilter}{\func{AugFilter}}
\newcommand{\augLeft}{\func{AugLeft}}
\newcommand{\augRange}{\func{AugRange}}
\newcommand{\mapReduce}{\func{MapReduce}}
\newcommand{\augProject}{\func{AugProject}}
\newcommand{\joinTwo}{\func{Join2}}
\newcommand{\filter}{\func{Filter}}
\newcommand{\forallnew}{\func{ForAll}}
\newcommand{\union}{\func{Union}}
\newcommand{\insertnew}{\func{Insert}}
\newcommand{\delete}{\func{Delete}}
\newcommand{\intersect}{\func{Intersect}}
\newcommand{\build}{\func{Build}}

\newcommand{\difference}{\func{Difference}}

\newcommand{\range}{\func{Range}}

\newcommand{\multiInsert}{{{\func{MultiInsert}}}}

\newcommand{\red}[1]{{\textcolor[rgb]{1.0,0.00,0.00}{#1}}}
\newcommand{\hide}[1]{} 

\newcommand{\para}[1]{\vspace{0.01in}\noindent\textbf{#1 }}

\newenvironment{newpart}{\par\color{blue}\begin{center}\textbf{- New! -}\end{center}}{\begin{center}\textbf{- End New! -}\end{center}\par}

\begin{document}

\title{PAM: Parallel Augmented Maps}


\author{
Yihan Sun\\
Carnegie Mellon University\\
yihans@cs.cmu.edu
\and
Daniel Ferizovic\\
Karlsruhe Institute of Technology\\
dani93.f@gmail.com
\and
Guy Blelloch\\
Carnegie Mellon University\\
guyb@cs.cmu.edu
}

\maketitle

\begin{abstract}
Ordered (key-value) maps are an important and widely-used data
type for large-scale data processing frameworks.  Beyond simple
search, insertion and deletion, more advanced operations such as range
extraction, filtering, and bulk updates form a critical
part of these frameworks.

We describe an interface for ordered maps that is augmented to
support fast range queries and sums, and introduce a parallel and
concurrent library called PAM (Parallel Augmented Maps) that
implements the interface.  The interface includes a wide variety of
functions on augmented maps ranging from basic insertion and deletion
to more interesting functions such as union, intersection, filtering,
extracting ranges, splitting, and range-sums.  We describe algorithms
for these functions that are efficient both in theory and practice.



As examples of the use of the interface and the performance of PAM we
apply the library to four applications: simple range sums, interval
trees, 2D range trees, and ranked word index searching.  The interface
greatly simplifies the implementation of these data structures over
direct implementations.  Sequentially the code achieves
performance that matches or exceeds existing libraries designed
specially for a single application, and in parallel our implementation
gets speedups ranging from 40 to 90 on 72 cores with 2-way
hyperthreading.
\end{abstract}

\section{Introduction}
\label{intro}

The \emph{map} data type (also called key-value store, dictionary,
table, or associative array) is one of the most important data types
in modern large-scale data analysis, as is indicated by systems such
as F1~\cite{shute13}, Flurry~\cite{flurry}, RocksDB~\cite{rocksdb},
Oracle NoSQL~\cite{oraclenosql}, LevelDB~\cite{leveldb}.  As such,
there has been significant interest in developing high-performance
parallel and concurrent algorithms and implementations of maps (e.g., see
Section~\ref{sec:related}).  Beyond simple insertion, deletion, and
search, this work has considered ``bulk'' functions over ordered maps,
such as unions~\cite{ding2011fast,inoue2014faster,ours}, bulk-insertion
and bulk-deletion~\cite{FS07,Barbuzzi10,WBTree}, and range
extraction~\cite{Brown2012,Prokopec12,Basin17}.

One particularly useful function is to take a ``sum''
over a range of
keys, where \emph{sum} means with respect to any associative
combine function (e.g., addition, maximum, or union).  As an example of such a \emph{range sum} consider a
database of sales receipts keeping the value of each sale ordered by
the time of sale.  When analyzing the data for certain trends, it is
likely useful to quickly query the sum or maximum of sales during a
period of time.  Although such sums can be implemented naively by
scanning and summing the values within the key range,
these queries can be answered much more efficiently using augmented
trees~\cite{preparata1985computational,CormenLeRi90}.  For example,
the sum of any range on a map of size $n$ can be answered in
$O(\log n)$ time.  This bound
is achieved by using a balanced binary tree and augmenting each node
with the sum of the subtree.

Such a data structure can also implement a significantly more general
form of queries efficiently.  In the sales receipts example they can
be used for reporting the sales above a threshold in $O(k\log (n/k+1))$ time
($k$ is the output size) if the augmentation is the maximum of sales, or in
$O(k + \log n)$ time~\cite{McCreight85} with a more complicated augmentation.
More generally they can be
used for interval queries, $k$-dimensional range queries, inverted
indices (all described later in the paper), segment intersection,
windowing queries, point location, rectangle intersection, range
overlaps, and many others.

\begin{table*}
\centering
\begin{tabular}{|c||c|c|@{}c@{}||@{}c@{}|r|r|r|@{}r@{ }|r|r|r|}
\hline
&\multicolumn{3}{c||}{\textbf{In Theory}} & \multicolumn{8}{c|}{\textbf{In Practice}} \\
 \multirow{2}{*}{\textbf{Application}} &\multicolumn{3}{c||}{\textbf{(Asymptotic bound)}} & \multicolumn{8}{c|}{\textbf{(Running Time in seconds)}} \\
\cline{2-12}
         &  \multicolumn{2}{c|}{\textbf{Construct}}            &      \multirow{2}{*}{\textbf{Query}}             &             \multicolumn{4}{c@{}|}{\textbf{Construct}}            &      \multicolumn{4}{@{}c|}{\textbf{Query}} \\
\cline{2-3}\cline{5-12}
           &       \textbf{Work} &      \textbf{Span} &            &                         \textbf{Size} &\multicolumn{1}{c|}{\textbf{Seq.}} & \multicolumn{1}{c|}{\textbf{Par.}} & \multicolumn{1}{c|}{\textbf{Spd.}} &    \textbf{Size} & \multicolumn{1}{c|}{\textbf{Seq.}} &\multicolumn{1}{c|}{\textbf{Par.}}&\multicolumn{1}{c|}{\textbf{Spd.}}       \\
\hline
\textbf{Range Sum} &  $O(n\log n)$ &   $O(\log n)$ &   $O(\log n)$ &$10^{10}$&   1844.38     &  28.24 &65.3 & $10^8$ &271.09&	 3.04	 & 89.2\\
\hline
\textbf{Interval Tree} &  $O(n\log n)$ &   $O(\log n)$ &   $O(\log n)$             &$10^8$&     14.35  &0.23	&	63.2 &$10^8$  &     53.35 &0.58& 92.7\\
\hline
\textbf{2d Range Tree} & $O(n \log n)$ & $O(\log^3 n)$ & $O(\log^2 n)$             &$10^8$&      197.47&	3.10&	63.7 &  $10^6$  & 48.13&	0.55	&87.5\\
\hline
\textbf{Inverted Index} & $O(n \log n)$ & $O(\log^2 n)$ &          *             &$1.96\times10^9$  &     1038 &     12.6 & 82.3& $10^5$ &   368 &4.74&77.6\\
\hline
\end{tabular}
\caption{The asymptotic cost and experimental results of the applications using PAM. \textbf{Seq.} = sequential, \textbf{Par.} =
  Parallel (on 72 cores with 144 hyperthreads), \textbf{Spd.} = Speedup. ``Work'' and ``Span'' are used to evaluate the theoretical bound of parallel algorithms (see Section \ref{sec:algorithm}). *: Depends on the query.\vspace{-.2in}}\label{tab:introtab}
\end{table*}

Although there are dozens of implementations of efficient range sums,
there has been very little work on parallel or concurrent
implementations---we know of none for the general case, and only two
for specific applications~\cite{agarwal16,kim13}.  In this paper we
present a general library called PAM (Parallel Augmented Maps) for
supporting in-memory parallel and concurrent maps with range sums.
PAM is implemented in C++.  We use \emph{augmented value} to refer to
the abstract ``sum'' on a map (defined in Section \ref{sec:augmap}).
When creating a map type the user specifies two augmenting functions
chosen based on their application: a \emph{base} function $g$ which gives the
augmented value of a single element, and a \emph{combine} function $f$ which
combines multiple augmented values, giving the augmented value of the
map.  The library can then make use of the functions to keep ``partial
sums'' (augmented values of sub-maps) in a tree that asymptotically
improve the performance of certain queries.

Augmented maps in PAM support standard functions on ordered maps
(which maintain the partial sums), as well as additional function
specific to augmented maps (see Figure~\ref{fig:augmap} for a partial list).
The standard functions include simple functions such as insertion, and
bulk functions such as union.  The functions specific to augmented
maps include efficient range-sums, and filtering based on augmented
values.
PAM uses theoretically efficient parallel algorithms for all bulk
functions, and is implemented based on using the ``join'' function to
support parallelism on balanced trees~\cite{ours}.  We extend the
approach of using joins to handle augmented values, and also give
algorithms based on ``join'' for some other operations such as
filtering, multi-insert, and mapReduce.

PAM uses functional data structures and hence the maps are fully
persistent---updates will not modify an existing map but will create
a new version~\cite{driscoll1986making}.  Persistence is useful in
various applications, including the range tree and inverted index
applications described in this paper.  It is also useful in supporting
a form of concurrency based on snapshot isolation.  In particular each
concurrent process can atomically read a snapshot of a map, and can
manipulate and modify their ``local'' copy without affecting the view
of other users, or being affected by any other concurrent modification to the
shared copy\footnote{Throughout the paper we use \emph{parallel} to
  indicate using multiple processors to work on a single bulk
  function, such as multi-insertion or filtering, and we use
  \emph{concurrent} to indicate independent ``users'' (or processes)
  asynchronously accessing the same structure at the same time.}.
However PAM does not directly support traditional concurrent updates
to a shared map.  Instead concurrent updates need to be batched and
applied in bulk in parallel.

We present examples of four use cases for PAM along with experimental
performance numbers.  Firstly we consider the simple case of
maintaining the sum of the values in a map using integer addition.
For this case we report both sequential and parallel times for a wide
variety of operations (union, search, multi-insert,
range-sum, insertion, deletion, filter, build).  We also
present performance comparisons to other implementations of maps that
do not support augmented values.  Secondly we use augmented maps to
implement interval trees.  An interval tree maintains a set of
intervals (e.g. the intervals of times in which users are logged into
a site, or the intervals of time for FTP connections) and can quickly
answer queries such as if a particular point is covered by any interval
(e.g. is there any user logged in at a given time).  Thirdly we
implement 2d range trees.  Such trees maintain a set of points in 2
dimensions and allow one to count or list all entries within a given
rectangular range (e.g. how many users are between 20 and 25 years old
and have salaries between \$50K and \$90K).  Such counting queries can
be answered in $O(\log^2 n)$ time.  We present performance comparison
to the sequential range-tree structure available in
CGAL~\cite{cgal:range}.  Finally we implement a weighted inverted
index that supports and/or queries, which can quickly return the top $k$
matches.  The theoretical cost and practical performance of these four
applications are shown in Table \ref{tab:introtab}.


The main contributions of this paper are:
\begin{enumerate}
\item An interface for \emph{augmented maps} (Section~\ref{sec:augmap}).
\item Efficient parallel algorithms and an implementation for the
  interface as part of the PAM library (Section~\ref{sec:algorithm}).
\item Four example applications of the interface (Section~\ref{sec:application}).
\item Experimental analysis of the examples (Section~\ref{sec:exp}).
\end{enumerate}

\section{Related Work}
\label{sec:related}

Many researchers have studied concurrent algorithms and
implementations of maps based on balanced search trees focusing on
insertion, deletion and
search~\cite{KL80,Lersen00,FS07,BCCO10,Barbuzzi10,WBTree,BE13,BER14,DVY14,Levandoski13,NM14}.
Motivated by applications in data analysis recently researchers have
considered mixing atomic range scanning with concurrent updates
(insertion and deletion)~\cite{Brown2012,Prokopec12,Basin17}.  None of
them, however, has considered sub-linear time range sums.

There has also been significant work on parallel algorithms and
implementations of bulk operations on ordered maps and sets~\cite{PVW83,PP01,BR98,FS07,ding2011fast,inoue2014faster,WBTree,ours,akhremtsevsanders}.
Union and intersection, for example, are available as part of the
multicore version of the C++ Standard Template Library~\cite{FS07}.
Again none of this work has considered fast range sums.
There has been some work on parallel data structures for specific applications
of range sums such as range trees~\cite{kim13}.

There are many theoretical results on efficient sequential
data-structures and algorithms for range-type queries using augmented
trees in the context of specific applications such as interval
queries, k-dimensional range sums, or segment intersection queries
(see e.g.~\cite{de2000computational}).  Several of these approaches
have been implemented as part of systems~\cite{Kriegel00,cgal:range}.
Our work is motivated by this work and our goal is to parallelize many
of these ideas and put them in a framework in which it is much easier
to develop efficient code.   We know of no other general framework
as described in Section~\ref{sec:augmap}.


Various forms of range queries have been considered in the context of
relational databases~\cite{Guttman84,Gray97,Ho97,Chan99,Gao05}.  Ho
et. al.~\cite{Ho97} specifically consider fast range sums.  However
the work is sequential, only applies to static data, and requires that
the sum function has an inverse (e.g. works for addition, but not
maximum).  More generally, we do not believe that traditional (flat)
relational databases are well suited for our approach since we use
arbitrary data types for augmentation---e.g. our 2d range tree has
augmented maps nested as their augmented values.  Recently there has
been interest in range queries in large clusters under systems such as
Hadoop~\cite{Kangaroo16,agarwal16}.  Although these systems can
extract ranges in work proportional to the number of elements in the
range (or close to it), they do not support fast range sums.  None of
the ``nosql'' systems based on key-value
stores~\cite{flurry,leveldb,rocksdb,oraclenosql} support fast range
sums.

\newcommand{\mto}{\rightarrow}
\newcommand{\mb}[1]{\mbox{\textbf{#1}}}
\newcommand{\intz}{\mathbb{Z}}
\newcommand{\mm}{\mathbb{AM}}
\newcommand{\opt}{\mb{option}}
\newcommand{\seq}{\mb{seq.}}
\newcommand{\bool}{\mb{bool}}
\begin{figure}
  \flushleft
  \textbf{(Partial) Interface AugMap} $\mm(K, V, A, <, g, f, I)$ :
  \[
  \begin{array}{lcl}
  \hline
    \mb{empty}~\mbox{or}~\emptyset & : & M\\
    \mb{size} & : & M \mto \intz \\
    \mb{single} & : & K \times V \to M\\
    \mb{find} & : & M \times K \mto V\cup\{\Box\} \\
    \mb{insert} & : & M \times K \times V \times (V\times V \mto V)\mto M\\
    \mb{union} & : & M \times M \times (V\times V\mto V) \mto M\\
    \mb{filter} & : & (K \times V \mto \bool) \times M \mto M\\
    \mb{upTo} & : & M\times K \mto M\\
    \mb{range} & : & M \times K \times K \mto M\\
    \mb{mapReduce} & : & (K \times V \mto B) \times (B \times B \mto B)
                      \times B\\
                        & & \times~ M \mto B\\
    \mb{build} & : & (K \times V)~\seq \times (V\times V\mto V)\mto M\\
    \hdashline
    \mb{augVal} & : & M \mto A\\
    \mb{augLeft} & : & M \times K \mto A\\
    \mb{augRange} & : & M \times K \times K \mto A\\
    \mb{augFilter} & : & (A \mto \bool) \times M \mto M\\
    \mb{augProject} & : & (A \mto B) \times (B \times B \mto B)\\
                 &   & \times~ M \times K \times K \mto B\\
\hline
  \end{array}
  \]\vspace{-.2in}
  \caption{The (partial) interface for an augmented map, with key type
    $K$, value type $V$, and augmented value type $A$.  The augmenting monoid
    is $(A,f,I)$.  Other functions not listed include \mb{delete}, \mb{intersect},
    \mb{difference}, \mb{split}, \mb{join}, \mb{downTo},
    \mb{previous}, \mb{next}, \mb{rank}, and \mb{select}. In the table \seq{} means a sequence.
    }
  \label{fig:augmap}\vspace{-.2in}
\end{figure}

\section{Augmented Maps}
\label{sec:augmap}
\emph{Augmented maps}, as defined here, are structures that associate
an ordered \emph{map} with a ``sum'' (the \emph{augmented value}) over all entries in the map.  It
is achieved by using a base function $g$ and a combine function $f$. More formally, an augmented map type $\mm$ is
parameterized on the following:\\
\begin{tabular}{lll}
$K,$ & key type\\
$<~: K \times K \mto \bool,$ & total ordering on the keys\\
$V,$ & value type\\[.07in]
$A,$ & augmented value type\\
$g : K \times V \mto A,$ & the base function\\
$f :A \times A \mto A,$ & the combine function\\
$I : A$ & identity for $f$
\end{tabular}\\[.05in]
The first three parameters correspond to a standard ordered map,
and the last four are for the augmentation.  $f$ must
be associative ($(A, f, I)$ is a monoid), and we use $f(a_1,a_2,\ldots,a_n)$ to mean any nesting.  Then the \emph{augmented value} of a map $m = \{(k_1,v_1),(k_2,v_2),\dots,(k_n,v_n)\}$ is defined as:
$$\mathcal{A}(m)=f(g(k_1,v_1),g(k_2,v_2),\dots,g(k_n,v_n))$$
As an example, the
augmented map type:
\begin{equation}
\label{eqn:exampleeqn}
\mm(\intz, <_{\intz}, \intz, \intz, (k,v) \mto v, +_{\intz}, 0)
\end{equation}
defines an augmented map with integer keys and values, ordered by
$<_{\intz}$, and for which the augmented value of any map of this type
is the sum of its values.

An augmented map
type supports an interface with standard functions on ordered maps as
well as a collection of functions that make use of $f$ and $g$.
Figure~\ref{fig:augmap} lists an example interface,
which is the one used in this paper and supported by PAM.
In the figure,
the definitions above the dashed line are standard
definitions for an ordered map.  For example, the $\mb{range}(m, k_1, k_2)$
extracts the part of the map between keys $k_1$ and $k_2$, inclusive,
returning a new map.
The $\mb{mapReduce}(g', f', I', m)$ applies the function $g'$ to each
element of the map $m$, and then sums them with the associative function
$f'$ with identity $I'$. Some functions listed in Figure \ref{fig:augmap}, such
as \union{}, \insertnew{} and \build{}, take an addition argument $h$,
which itself is a function. When applicable it combines values of all entries with
the same key.
For example the $\mb{union}(m_1,m_2,h)$ takes union of two maps,
and if any key appears in both maps, it combines their values using $h$.

Most important to this paper are the definitions below the dashed line, which
are functions specific to augmented maps.  All of them can be computed
using the functions above the dashed line.  However, they can be much more
efficient by maintaining the augmented values of sub-maps (partial
sums) in a tree structure.  Table~\ref{tab:costs} gives the asymptotic
costs of the functions based on the implementation described in
Section \ref{sec:algorithm}, which uses augmented balanced search
trees.  The function $\mb{augVal}(m)$ returns $\mathcal{A}(m)$, which is equivalent to $\mb{mapReduce}(g,f,I,m)$ but can run in
constant instead of linear work.  This is because the functions
$f$ and $g$ are chosen ahead of time and integrated into the augmented
map data type, and therefore the sum can be maintained during updates.
The function $\mb{augRange}(m,k_1,k_2)$ is equivalent to
$\mb{augVal}(\mb{range}(m,k_1,k_2))$ and $\mb{augLeft}(m,k)$ is
equivalent to $\mb{augVal}(\mb{upTo}(m,k))$. These can also be implemented
efficiently using the partial sums.


The last two functions accelerate two common queries on augmented maps,
but are only applicable
when their function arguments meet certain requirements.  They also
can be computed using the plain map functions, but can be much more
efficient when applicable.
The \mb{augFilter}$(h,m)$ function is equivalent to
$\mb{filter}(h', m)$, where $h':K\times V\mapsto \bool$ satisfies
$h(g(k,v))\Leftrightarrow h'(k,v)$.
It is only applicable
if $h(a) \vee h(b) \Leftrightarrow h(f(a,b))$
for any $a$ and $b$ ($\vee$ is the logical or).
In this case the $\mb{filter}$ function can
make use of the partial sums.
For example, assume the values in the map are boolean values, $f$ is a logical-or, $g(k,v)=v$, and we want
to filter the map using function $h'(k,v)=v$. 
In this case we can filter out a whole sub-map once we see it
has \ctext{false} as a partial sum. Hence we can set $h(a)=a$ and directly use \mb{augFilter}$(m,h)$.
\hide{
In our example of
sales receipts over time with augmentation to be the maximum sale, this function can be used to filter
out all sales above a given amount very much more efficiently than scanning
a whole range of sales.  }
The function is used in interval trees
(Section~\ref{sec:interval}).  The
\mb{augProject}$(g', f',m,k_1,k_2)$ function is equivalent to
$g'(\mb{augRange}(m,k_1,k_2))$.  It requires, however, that $(B,f',g'(I))$ is a
monoid and that $f'(g'(a),g'(b)) = g'(f(a,b))$.  This function is useful
when the augmented values are themselves maps or other large data
structures.  It allows projecting the augmented values down onto another type by $g'$ (e.g.
project augmented values with complicated structures to values like their sizes)
then summing them by $f'$, and is much more efficient when applicable. For example in range trees where
each augmented value is itself an augmented map, it greatly improves performance for queries.

\newcommand{\myparagraph}[1]{\vspace{-.06in}\paragraph{#1}}

\section{Data Structure and Algorithms}
\label{sec:algorithm}
In this section we outline a data structure and associated
algorithms used in PAM that can be used to efficiently implement augmented maps.
Our data structure is based on abstracting the balancing criteria of a
class of trees (e.g. AVL or Red-Black trees) in terms of a single
\join{} function~\cite{ours}, which joins two maps with a key between
them (defined more formally below).  Prior work describes parallel
algorithms for \union{}, \difference{}, and \intersect{} for standard
un-augmented sets and maps using just \join~\cite{ours}.  Here we
extend the methodology to handle augmentation, and also describe
some other functions based on join.
Because the balancing criteria is fully abstracted in
\join{}, similar algorithm can be applied to AVL trees~\cite{avl}
red-black trees~\cite{redblack},
weight-balanced trees and treaps~\cite{SA96}. We implemented all of them in PAM.
By default, we use weight-balanced
trees~\cite{weightbalanced} in PAM, because it does not require extra balancing criteria
in each node (the node size is already stored), but users can change to any specific balancing scheme
using C++ templates.

\myparagraph{Augmentation.}
We implement augmentation by storing with every tree node the
augmented sum of the subtree rooted at that node.  This localizes
application of the augmentation functions $f$ and $g$ to when a node
is created or updated\footnote{For supporting persistence, updating a
tree node, e.g., when it is involved in rotations and is set a new child,
usually results in the creation of a new node. See details in the persistence part.}.
In particular when creating a node with a left child $L$,
right child $R$, key $k$ and value $v$ the augmented value can be
calculated as $f(\mathcal{A}(L), f(g(k,v), \mathcal{A}(R)))$, where $\mathcal{A}(\cdot)$ extracts the
augmented value from a node.  Note that it takes two
applications of $f$ since we have to combine three values, the left,
middle and right.  We do not store $g(k,v)$.  In our algorithms,
creation of new nodes is handled in \join{}, which also deals with
rebalancing when needed.  Therefore all the algorithms and code that
do not explicitly need the augmented value are unaffected by and even oblivious of
augmentation.


\myparagraph{Parallelism.}
We use fork-join parallelism to implement internal parallelism for
the bulk operations. In pseudocode the notation ``$s_1~||~s_2$''
means that the two statements $s_1$ and $s_2$ can run in parallel, and
when both are finished the overall statement finishes.  In most cases
parallelism is over the structure of the trees---i.e. applying some
function in parallel over the two children, and applying this
parallelism in a nested fashion recursively (Figure~\ref{alg:pseudocode}
shows several examples).  The only exception is \mb{build}, where we
use parallelism in a sort and in removing duplicates.  In the PAM
library fork-join parallelism is implemented with the cilkplus
extensions to C++~\cite{Cilk}.  We have a granularity set so
parallelism is not used on very small trees.
\renewcommand{\arraystretch}{1.2}
\begin{table}[!h!t]
\begin{center}
	\begin{tabular}{l@{  }|@{  }c@{  }|@{  }c}
	\hline
	\textit{\textbf{Function}} & \textit{\textbf{Work}} & \textit{\textbf{Span}} \\
	\hline
\multicolumn{3}{l}{\textbf{Map operations}}\\
\hline
	insert, delete, find, first, & \multirow{3}{*}{$\log n$} & \multirow{3}{*}{$\log n$}\\
last, previous, next, rank,&&\\
select, upTo, downTo&&\\
\hline
\multicolumn{3}{l}{\textbf{Bulk operations}}\\
\hline
join & $\log n -\log m$ &$~\log n -\log m$\\
\hline
	union$^{*}$, intersect$^{*}$,  &  \multirow{2}{*}{~$m \log \left(\frac{n}{m} + 1\right)$} & \multirow{2}{*}{$\log n \log m$} \\
difference$^{*}$ & \\
	\hline
	mapReduce & $n$ & $\log n$ \\
	\hline
	filter & $n$ & $\log^2 n$ \\
	\hline
	range, split, join2 & $\log n$ & $\log n$ \\
	\hline
	build & $n \log n$ & $\log n$ \\
	\hline
\multicolumn{3}{l}{\textbf{Augmented operations}}\\
	\hline
augVal & 1 & 1\\
\hline
augRange, augProject  & $\log n$& $\log n$\\
\hline
augFilter (output size $k$) &~$k\log (n/k+1)$ & $\log^2 n$\\
\hline
	\end{tabular}
	\end{center}
\caption{The core functions in PAM and their
      asymptotic costs (all big-O).  The cost is given under the assumption that
      the base function $g$, the combine function $f$ and the functions as parameters (e.g., for \augProject{}) take constant time to return. For the functions noted with $^*$, the efficient algorithms with bounds shown in the table are introduced and proved in~\cite{ours}. For functions with two input maps (e.g., \union{}), $n$ is the size of the larger input, and $m$ of
      the smaller.}
    \label{tab:costs}\vspace{-.2in}
\end{table}

\myparagraph{Theoretical bounds.}  To analyze the asymptotic costs in
theory we use work ($W$) and span ($S$), where work is the total
number of operations (the sequential cost) and span is the length of
the critical path~\cite{CormenLeRi90}.  Almost all the algorithms we
describe, and implemented in PAM, are asymptotically optimal in
terms of work in the comparison model, i.e., the total number of comparisons.
Furthermore
they achieve very good parallelism (i.e. polylogarithmic span).  Table
\ref{tab:costs} list the cost of some of the functions in PAM.  
When the augmenting functions $f$ and $g$ both take constant
time, the augmentation only affects all the bounds by a constant
factor.  Furthermore experiments show that this constant factor is
reasonably small, typically around 10\% for simple functions such as
summing the values or taking the maximum.

\begin{figure}
  \input{mlcode}
  \begin{lstlisting}
Union$(T_1, T_2, h)$ =
  if $T_1=\emptyset$ then $T_2$
  else if $T_2=\emptyset$ then $T_1$
  else let $\langle L_2, k, v, R_2 \rangle = T_2 $
            and $\langle L_1, v', R_2\rangle = $ Split$(T_1, k)$
            and $L =$ Union$(L_1, L_2)$ || $R=$ Union$(R_1, R_2)$
        in if $v'\ne \Box$ then Join$(L,k,h(v,v'),R)$
            else Join$(L,k,v,R)$@\vspace{.05in}@
Insert$(T,k,v,h)$ =
  if $T = \emptyset$ then Singleton$(k,v)$
  else let $\langle L, k', v', R \rangle = T $ in
     if $k < k'$ then Join(Insert$(L, k, v), k', v', R$)
     else if $k > k'$ then Join$(L, k', v',$ Insert$(R, k, v))$
     else Join$(L,k,h(v',v),R)$@\vspace{.05in}@
MapReduce$(T, g',f',I')$ =
  if $T=\emptyset$ then $I'$
  else let $\langle L, k, v, R \rangle = T $
            and $L' = $ MapReduce$(L,g',f',I')$ || 
                $R' =$ MapReduce$(R,g',f',I')$
         in $f'(L',g'(k,v),R')$@\vspace{.05in}@
AugLeft$(T, k')$ =
  if $T=\emptyset$ then $I$
  else let $\langle L, k, v, R \rangle = T$
         in if $k'<k$ then AugLeft$(L,k')$
             else $f(\mathcal{A}(L),g(k,v),$AugLeft$(R,k'))$  @\vspace{.05in}@
Filter$(T, h)$ =
  if $T=\emptyset$ then $\emptyset$
  else let $\langle L, k, v, R \rangle = T $
            and $L' = $ Filter$(L,h)$ || $R' =$ Filter$(R,h)$
        in if $h(k,v)$ then Join$(L',k,v,R')$ else Join2$(L',R')$@\vspace{.05in}@
AugFilter$(T, h)$ =
  if $(T=\emptyset)$ or $(\neg h(\mathcal{A}(T)))$ then $\emptyset$
  else let $\langle L, k, v, R \rangle = T $
            and $L' = $ AugFilter$(L,h)$ ||
                   $R' =$ AugFilter$(R,h)$
        in if $h(g(k,v))$ then Join$(L',k,v,R')$ else Join2$(L',R')$@\vspace{.05in}@
Build'$(S,i,j)$ =
  if $i = j$ then $\emptyset$
  else if $i+1=j$ then Singleton$(S[i])$
  else let $m = (i + j) / 2$
            and $L = $ Build'$(S,i,m)$ || $R =$ Build'$(S,m+1,j)$
         in Join$(L,S[m],R)$@\vspace{.05in}@
Build$(S)$ =
   Build'(RemoveDuplicates(Sort$(S)$)$,0,|S|)$
\end{lstlisting}
\vspace{-.15in}
  \caption{\small Pseudocode for some of the functions on augmented maps.
    \union{} is from~\cite{ours}, the rest are new. For an associated binary function $f$,
    $f(a,b,c)$ means $f(a,f(b,c))$.}
  \label{alg:pseudocode}
\end{figure}

\myparagraph{Join, Split, Join2 and Union.}
\label{sec:join}
As mentioned, we adopt and extend the methodology introduced
in~\cite{ours}, which builds all map functions using \join$(L,k,R)$.
The \join{} function takes two ordered maps $L$ and $R$ and a
key-value pair $(k,v)$ that sits between them (i.e. $\max(L) < k <
\min(R)$) and returns the composition of $L$, $(k,v)$ and $R$.  Using
\join{} it is easy to implement two other useful functions: \split{}
and \joinTwo{}.  The function $\langle L, v, R \rangle=$
\func{Split}$(A,k)$ splits the map $A$ into the elements less than $k$
(placed in $L$) and those greater (placed in $R$).  If $k$ appears in
$A$ its associated value is returned as $v$, otherwise $v$ is set to be an empty value noted as $\Box$.  The function
$T=$\func{Join2}$(T_L, T_R)$ works similar to \join{}, but without
the middle element.  These are useful in the other algorithms.
Algorithmic details can be found in \cite{ours}.

The first function in Figure~\ref{alg:pseudocode} defines an algorithm
for \union{} based on \split{} and \join{}.  We add a feature that
it can accept a function $h$ as parameter.
If one key appears in both sets, the values are combined using $h$.
In the pseudocode we
use $\langle L, k, v, R \rangle = T$ to extract the left child, key,
value and right child from the tree node $T$, respectively.  The pseudocode is
written in a functional (non side-effecting) style.  This matches our
implementation, as discussed in \textbf{\em persistence} below.

\myparagraph{Insert and Delete.}
Instead of the classic implementations of \insertnew{} and \delete{},
which are specific to the balancing scheme, we define versions based
purely on \join{}, and hence independent of the balancing scheme. Like the
\union{} function, \insertnew{} also takes an addition function $h$ as input, 
such that if the key to be inserted is found in the map, the values will be combined 
by $h$. The
algorithm for insert is given in Figure~\ref{alg:pseudocode}, and the
algorithm for deletion is similar.
The algorithms run in $O(\log n)$ work (and span since sequential).
One might expect that abstracting insertion using
\join{} instead of specializing for a particular balance criteria has
significant overhead.    Our experiments show this is not the case---and
even though we maintain the reference counter for persistence, we are only 17\% slower
sequentially than the highly-optimized C++ STL library (see section~\ref{sec:exp}).

\myparagraph{\mb{Build}.}
\hide{An augmented map can be created from a sequence of key-value pairs
using a balanced divide-and-conquer over the input sequence and
combining with union.  If the augmentation functions $f$ and $g$ take
constant time then this will take $O(n \log n)$ work and $O(\log^3 n)$
span.}  To construct an augmented map from a sequence of key-value pairs
we first sort the sequence by the keys, then
remove the duplicates (which are contiguous in sorted order), and
finally use a balanced divide-and-conquer with \join{}.    The algorithm is given in Figure~\ref{alg:pseudocode}.
The work is then $O(W_{\mbox{\small{sort}}(n)} + W_{\mbox{\small{remove}}(n)} + n)$ and
the span is $O(S_{\mbox{\small{sort}}(n)} + S_{\mbox{\small{remove}}(n)} + \log n)$.
For work-efficient sort and remove-duplicates with $O(\log n)$ span
this gives the bounds in Table~\ref{tab:costs}.

\myparagraph{Reporting Augmented Values.}
\label{sec:augfunctions}
As an example, we give the algorithm of \augLeft{}$(T,k')$ in Figure \ref{alg:pseudocode}, which returns the augmented value of all entries with keys less than $k'$. It compares the root of $T$ with $k'$, and if $k'$ is smaller, it calls \augLeft{} on its left subtree. Otherwise the whole left subtree and the root should be counted. Thus we directly extract the augmented value of the left subtree, convert the entry in the root to an augmented value by $g$, and recursively call \augLeft{} on its right subtree. The three results are combined using $f$ as the final answer. This function visits at most $O(\log n)$ nodes, so it costs $O(\log n)$ work and span assuming $f$, $g$ and $I$ return in constant time. The \augRange{} function, which reports the augmented value of all entries in a range, can be implemented similarly with the same asymptotical bound.

\myparagraph{Filter and AugFilter.}
\label{sec:augfunctions}
The filter and augFilter function both select all entries in the map
satisfying condition $h$.  For a (non-empty) tree \texttt{T},
\filter{} recursively filters its left and right branches in parallel,
and combines the two results with \join{} or \joinTwo{} depending on
whether $h$ is satisfied for the entry at the root.  It takes
linear work and $O(\log^2 n)$ span for a balanced tree.
The \mb{augFilter}$(h,m)$ function has the
same effect as $\mb{filter}(h', m)$, where $h':K\times V\mapsto \bool$ satisfies
$h(g(k,v))\Leftrightarrow h'(k,v)$ and is only applicable if $h(a)\vee
h(b)\Leftrightarrow h(f(a,b))$.  This can asymptotically improve
efficiency since if $h(\mathcal{A}(T))$ is false, then we know that
$h$ will not hold for any entries in the tree $T$\footnote{Similar methodology
can be applied if there exists a function $h''$ to decide if all entries in a subtree
will be selected just by reading the augmented value. }, so the search can
be pruned (see Figure \ref{fig:augmap}).  For $k$ output entries, this function
takes $O(k\log(n/k+1))$
work, which is asymptotically smaller than $n$ and is significantly more
efficient when $k$ is small. Its span is $O(\log^2 n)$.

\myparagraph{Other Functions.}
We implement many other functions in PAM, including all in Figure \ref{fig:augmap}. In \mapReduce{}$(g',f',I',T)$,  for example, it is recursively applied on $T$'s two subtrees in parallel, and $g'$ is applied to the root. The three values are then combined using $f'$. The function \augProject{}$(g',f',m,k_1,k_2)$ on the top level adopts a similar method as \augRange{} to get related entries or subtrees in range $[k_1,k_2]$, projects $g'$ to their augmented values and combine results by $f'$.

\hide{
\myparagraph{Other Functions.}
\label{sec:filter}
We also implemented all the functions listed in Figure \ref{fig:augmap}.
Since the other algorithms are rather straight-forward we
do not give discussions here. As an example, we show the \texttt{filter} in Algorithm \ref{algo:filter}.
In addition to the plain algorithm,
we apply extra optimizations to improve parallelism. We give the C++ code for \texttt{filter} in Figure \ref{fig:filter}.
For a (non-empty) node \texttt{b}, \texttt{filter}
recursively filters its left and right branch, but these recursive calls
are made parallel only when the size of \texttt{b} is larger than
\texttt{cutoff} (line~\ref{line:cutoff}).
Due to the nested parallelism in the recursive calls there is
significant parallelism (typically much more than the number of cores
on the machine). The \texttt{decrement} is part of
the reference counting collector, and the \texttt{copy\_if\_needed} is
part of an optimization for persistence described below.
}

\myparagraph{Persistence.}
The PAM library uses functional data
structures, and hence does not modify
existing tree nodes but instead creates new ones~\cite{okasaki1999purely}.  This is not
necessary for implementing augmented maps, but is helpful in
the parallel and concurrent implementation.  Furthermore the fact that
functional data structures are persistent (no existing data is
modified) has many applications in developing efficient data
structures~\cite{driscoll1986making}.  In this paper three of our four
applications (maintaining inverted indices,
interval trees and range trees) use persistence in a critical way.  The \join{}
function copies nodes along the join path, leaving the two original
trees unchanged.  All of our code is built on \join{} in the
functional style, returning new trees rather than modifying old ones.
Such functional data structures mean that parts of trees are shared,
and that old trees for which there are no longer any pointers need to
be garbage collected.  We use a reference counting garbage collector.
When the reference count is one we use a standard reuse
optimization---reusing the current node instead of collecting it
and allocating a new one~\cite{Hudak86}.

\myparagraph{Concurrency.}
In PAM any number of users can concurrently access and update their
local copy (snapshot) of any map.  This is relatively easy to
implement based on functional data structures (persistent) since each
process only makes copies of new data instead of modifying shared
data.  The one tricky part is maintaining the reference counts and
implementing the memory allocator and garbage collector to be safe for
concurrency.  This is all implemented in a lock-free fashion.  A
compare-and-swap is used for updating reference counts, and
a combination of local pools and a global pool are used for memory
allocation.  Updates to the shared instance of a map can be made
atomically by swapping in a new pointer (e.g., with a
compare-and-swap).  This means that updates are sequentialized.
However they can be accumulated and applied when needed in bulk using
the parallel multi-insert or multi-delete.

\section{Applications}
\label{sec:application}
In this section we describe applications that can be implemented using the PAM interface.
Our first application is given in Equation \ref{eqn:exampleeqn}, which is a map storing
integer keys and values, and keeping track of sum over values. In this section we
give three more involved applications of augmented
maps: interval trees, range trees and word indices (also called inverted indices).
We note that although we use trees as the implementation, the abstraction of the applications to
augmented maps is independent of representations. 

\hide{
\subsection{Using the PAM Interface}
We first briefly introduce the interfaces of PAM and show how to use them.
The library uses the \texttt{aug\_map} as the only interface and defines plain maps and sets as special cases of it.
To use it, users need to define the key, value and augmented value type as well as the augmented parameters $f$, $g$ and $a_{\emptyset}$ by C++ templates.
The interface \texttt{aug\_map<K, V, Aug>} requires the following templates:
\begin{itemize}
\item \textbf{\texttt{K}} -- The key type. This type must support a comparison function ($<$) that properly defines a total order over its elements.
\item \textbf{\texttt{V}} -- The value type.
\item \textbf{\texttt{Aug}} -- The augmentation type. It must contains the following attributes and methods:
\begin{itemize}
  \item \textbf{Typename \texttt{aug\_t}} is the augmented value type.
  \item \textbf{Static method \texttt{base(K, V)}} defines the augmented base function $g:K\times V \mapsto \augtype$.
  \item \textbf{Static method \texttt{combine(aug\_t, aug\_t)}} defines the augmented combine function $f:\augtype \times \augtype \mapsto \augtype$.
  \item \textbf{Static method \texttt{identity()}} returns $a_{\emptyset}$.
\end{itemize}
\end{itemize}
With the interface, implementing all the following applications is just as simple as properly defining the templates above with just a few lines of code.}

\subsection{Interval Trees}
\label{sec:interval}
We give an example of how to use our interface for interval
trees~\cite{Edelsbrunner80,preparata1985computational,CormenLeRi90,de2000computational,Kriegel00,IntervalCSharp}.
This data structure maintains a set of intervals on the real line,
each defined by a left and right endpoint.  Various queries can be
answered, such as a stabbing query which given a point reports whether it is in an interval.

\begin{figure}
  \centering
  \includegraphics[width=1.0\columnwidth]{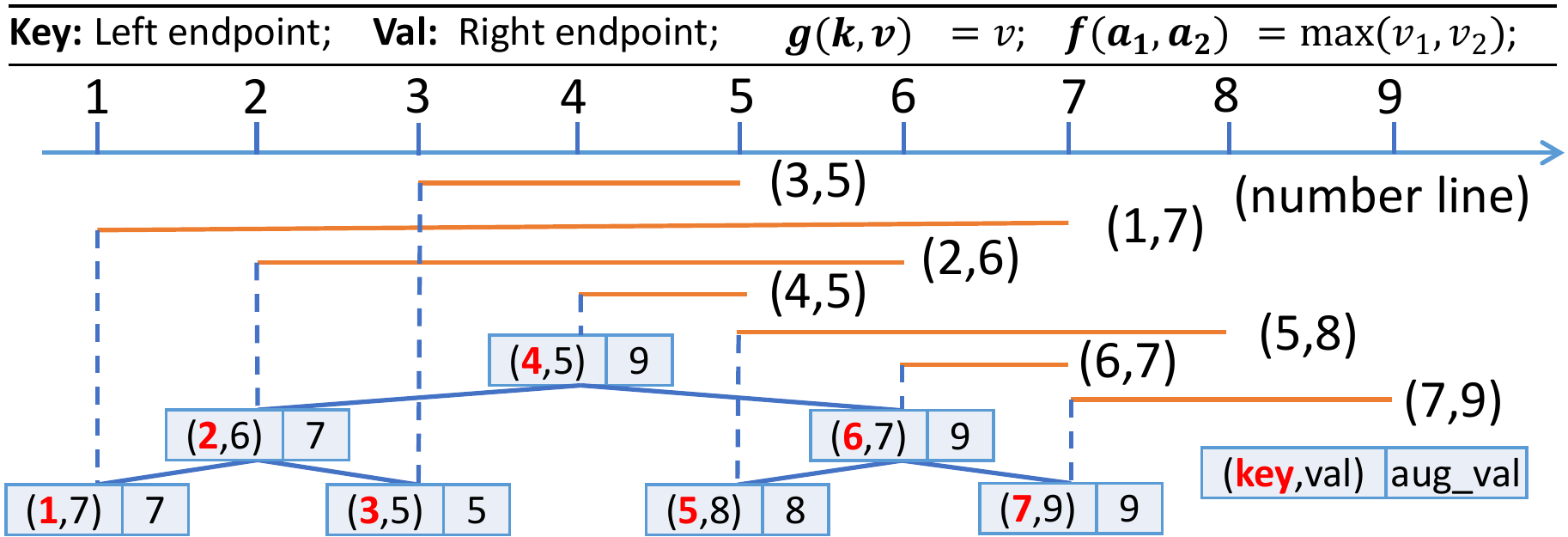}\\
  \caption{An example of an interval tree.}\label{fig:interval_illustration}
  \vspace{-.2in}
\end{figure}

There are various versions of interval trees. Here we discuss the version as
described in~\cite{CormenLeRi90}. In this version each interval is stored in a tree node,
sorted by the left endpoint (key).  A point $x$ is covered by in an
interval in the tree if the maximum right endpoint for all intervals
with keys less than $x$ is greater than $x$ (i.e. an interval starts
to the left and finishes to the right of $x$).  By storing at each
tree node the maximum right endpoint among all intervals in its
subtree, the stabbing query can be answered in $O(\log n)$ time.
An example is shown in Figure \ref{fig:interval_illustration}.

In our framework this can easily be implemented by using the left
endpoints as keys, the right endpoints as values, and using $\max$ as
the combining function.  The definition is:

\begin{tabular}{l}
$I=\mm(\mathbb{R},<_{\mathbb{R}},\mathbb{R},\mathbb{R},(k,v)\mapsto v, {\max}_{\mathbb{R}}, -\infty)$
\end{tabular}

Figure \ref{fig:interval_tree} shows the C++ code of the interval tree structure using PAM.  The entry with augmentation is
defined in \ctext{entry} starting from line~\ref{line:entry}, containing the key type \ctext{key\_t}, value type \ctext{val\_t}, comparison function \ctext{comp}, augmented value type (\ctext{aug\_t}), the base function $g$ (\ctext{base}),  the combine function $f$ (\ctext{combine}), and the identity of $f$ (\ctext{identity}). An
augmented map (line~\ref{line:amap}) is then declared as the interval tree structure with \ctext{entry}.
The constructor on line~\ref{line:imap} builds an interval tree from an array of $n$ intervals by directly calling the
  augmented-map constructor in PAM ($O(n \log n)$ work).
The function \ctext{stab(p)} returns if $p$ is
inside any interval using \ctext{amap::aug\_left(m,p)}. As defined in Section \ref{sec:augmap} and \ref{sec:algorithm},
this function returns the augmented sum, which is the $\max$ on values, of all entries with keys less than $p$.
As mentioned we need only to compare it with $p$.
The function \ctext{report\_all(p)} returns all
intervals containing $p$, which are those with keys
less than $p$ but values larger than $p$. We first get the sub-map in \ctext{m}
with keys less then $p$ (\ctext{amap::upTo(m,p)}), and filter all with values larger than $p$.
Note that $h(a)=(a>p)$ and the combine function $f(a,b)=\max(a,b)$ satisfy $h(a)\vee h(b)\Leftrightarrow h(f(a,b))$.
This means that to get all nodes with values $>p$, if the maximum value of a subtree is less than $p$,
the whole subtree can be discarded.
Thus we can apply \ctext{amap::aug\_filter} ($O(k \log (n/k+1))$ work for $k$
results),
which is more efficient than a plain filter.

\lstset{numbers=left}
\makeatletter%
\def\lst@PlaceNumber{\makebox[\dimexpr .3em+\lst@numbersep][l]{\normalfont
  \lst@numberstyle{\thelstnumber}}}%
\makeatother%

\begin{figure}[!h!t]
{\ttfamily\small
\begin{lstlisting}[language=C++,frame=lines,escapechar=@]
struct interval_map {
  using interval = pair<point, point>;
  struct entry {@\label{line:entry}@
    using key_t = point;
    using val_t = point;
    using aug_t = point;
    static bool comp(key_t a, key_t b)
      { return a < b;}
    static aug_t identity()
      { return 0;}
    static aug_t base(key_t k, val_t v)
      { return v;}
    static aug_t combine(aug_t a, aug_t b) {
      return (a > b) ? a : b;}
  };
  using amap = aug_map<entry>;@\label{line:amap}@
  amap m;@\cspace@
  interval_map(interval* A, size_t n) {@\label{line:imap}@
    m = amap(A,A+n); }@\cspace@
  bool stab(point p) {
    return (amap::aug_left(m,p) > p);}@\cspace@
  amap report_all(point p) {
    amap t = amap::up_to(m,p);
    auto h = [] (P a) -> bool {return a>p;}
    return amap::augFilter(t,h);};
\end{lstlisting}}
\vspace{-.2in}
\caption{The definition of interval maps using PAM in C++.}
\label{fig:interval_tree}
\end{figure}

\hide{
\begin{figure}[!h!t]
{\ttfamily\small
\begin{lstlisting}[language=C++,frame=lines,escapechar=@]
struct interval_map {
  using P = int;
  using interval = pair<P, P>;@\cspace@
  struct aug {@\label{line:aug}@
    using aug_t = P;
    static aug_t identity() {return -inf;}
    static aug_t base(P k, P v) {return v;}
    static aug_t combine(aug_t a, aug_t b) {
      return max(a,b);}  };@\cspace@
  auto less= [] (P a, P b) {return a<b;}
  using i_tree = aug_map<P,P,aug,less>;@\label{line:amap}@
  i_tree m;@\cspace@
  interval_map(interval* A, int n) {@\label{line:imap}@
    m.build(A,A+n); }@\cspace@
  bool stab(P p) {return (m.aug_left(p)>p);}@\cspace@
  vector<interval> report_all(point p) {
    vector<interval> vec;
    i_tree t = m.up_to(p);
    auto h = [] (P a) -> bool {return a>p;}
    return toSeq(t.augFilter(h));}  };
\end{lstlisting}}
\vspace{-.2in}
\caption{The definition of interval maps using PAM in C++.}
\label{fig:interval_tree}
\end{figure}}

\hide{
    while (I.second > p) {
      vec.push_back(I);
      a.remove(I.first);
      I = a.aug_left(p); }
        void insert(interval i) {m.insert(i);}@\cspace@}

\subsection{Range Trees}
\label{sec:rangetreeapp}
Given a set of $n$ points $\{p_i=(x_i, y_i)\}$ in the
plane, where $x_i\in X, y_i\in Y$, each point with weight $w_i\in W$, a \emph{2D range sum query} asks for the sum of
weights of points within a rectangle defined by a horizontal range
$(x_L, x_R)$ and vertical range $(y_L, y_R)$.
A \emph{2D range query} reports all points in the query window.
In this section, we describe how to adapt 2D range trees to the
PAM framework to efficiently support these queries.

The standard 2D range tree \cite{bentley1979decomposable,de2000computational,samet1990design}
 is a two-level tree (or map) structure.  The outer
level stores all the points ordered by
the x-coordinates. Each tree node stores
an inner tree with all points in its
subtree but ordered by the y-coordinates.  Then a range query can be done by
two nested queries on x- and y-coordinates respectively.
Sequential construction time and query time is $O(n\log
n)$ and $O(\log^2 n)$ respectively (the query time can be
reduced to $O(\log n)$ with reasonably complicated approaches).
\hide{\[R_I = \mm(X\times Y, W, <_Y, W, (k,v) \mto v, +_W, 0_W)\]
\[R_O = \mm(X\times Y, W, <_X, R_I, R_I.\text{single}, \cup, \emptyset) .\]
}

\begin{figure}
  \centering
  \includegraphics[width=0.95\columnwidth]{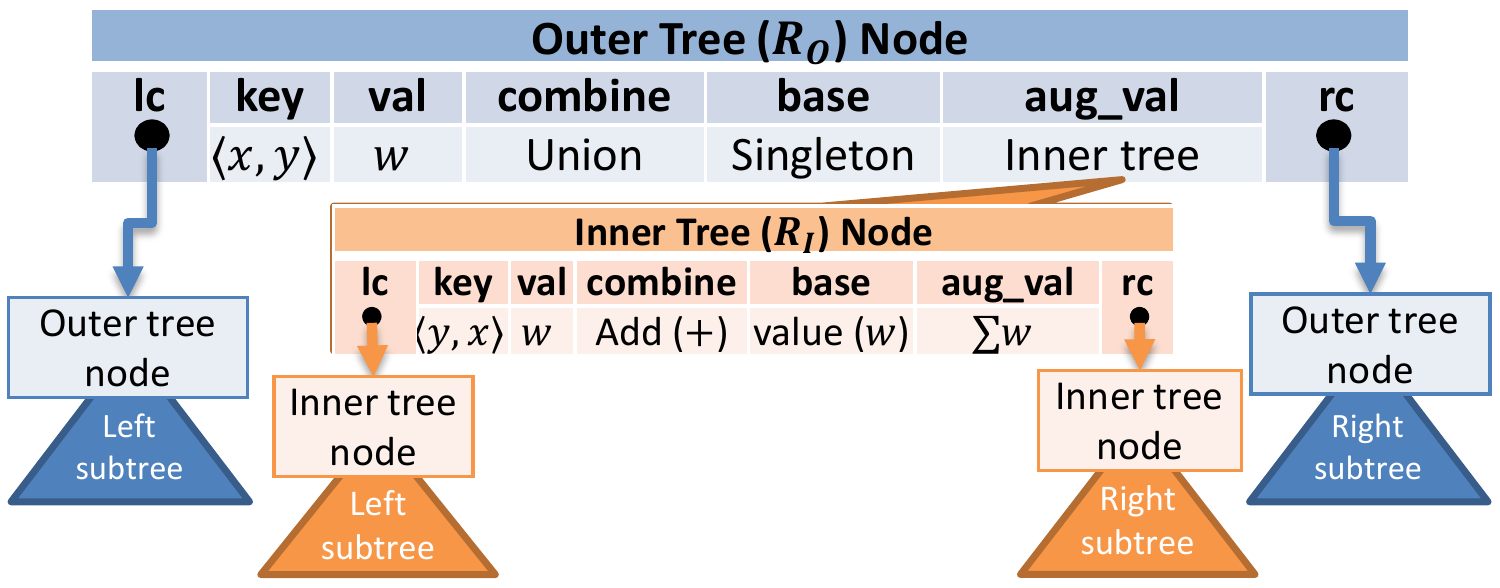}\\
  \caption{The range tree data structure under \ourname{} framework. In the illustration we omit some attributes such as size, reference counter and the \texttt{identity} function. Note that the functions \texttt{combine}, \texttt{base} and \texttt{identity} of both the outer tree and inner tree are static functions, so these functions shown in this figure actually do not take any real spaces.}\label{fig:rangetree_graph}
\end{figure}

In our interface the outer tree ($R_O$) is represented as an augmented
map in which keys are points (sorted by x-coordinates) and values are
weights.  The augmented value, which is the inner tree, is another
augmented map ($R_I$) storing all points in its subtree using the points as the key (sorted by
y-coordinates) and the weights as the value. The inner map is augmented by the sum
of the weights for efficiently answering range sums.  Union is used as the combine function for the outer
map.  The range tree layout is illustrated in in Figure
\ref{fig:rangetree_graph}, and the definition in our framework is:

{\small
\begin{tabular}{l@{}@{ }l@{ }@{}l@{}@{ }l@{ }@{}l@{ }@{}l@{ }@{}l@{ }@{}l@{ }@{}l@{ }@{}l@{ }@{}l@{ }@{}l@{ }}
$R_I$ &$=$& $\mm$&(&$P$, &$<_Y$, &$W$, &$W$, &$(k,v) \mto v$, &$+_W$, &$0_W$&)\\
$R_O$ &$=$& $\mm$&(&$P$, &$<_X$, &$W$, &$R_I$, &$R_I.$\text{singleton}, &$\cup$, &$\emptyset$&)
\end{tabular}}

Here $P=X\times Y$ is the point type. $W$ is the weight type. $+_W$ and $0_W$ are the addition function on $W$ and its  identity respectively.

It is worth noting that because of the persistence supported by the
PAM library, the combine function \union{} does not affect the inner
trees in its two children, but builds a new version of $R_I$
containing all the elements in its subtree.  This is important in
guaranteeing the correctness of the algorithm.

\hide{
\begin{figure}[!h!t]
\vspace{-0.1in}
{\ttfamily\small
\begin{lstlisting}[language=C++,frame=lines,escapechar=@]
typedef C int; typedef W double;
struct RangeQuery {
  using pnt = pair<C, C>;
  auto add=[](W w1, W w2) -> W {
    return w1+w2;}
  struct a_add {
    using aug_t = W;
    static aug_t base(pnt k, W v) {
      return v;}
    static aug_t combine(aug_t a, aug_t b) {
      return a+b; }
    static aug_t identity { return 0; } };
  using sec_tree = aug_map<pnt, W, a_add>;@\vspace{.03in}@
  struct a_union {
    using aug_t = sec_tree;
    static aug_t base(pnt k, W v) {
      return aug_t(make_pair(
        make_pair(k.second, k.first),v)); }
    static aug_t combine(aug_t a, aug_t b) {
      return map_union(a, b, add); }
    static aug_t identity { return aug_t(); } };@\vspace{.03in}@
  aug_map<pnt, W, a_union> range_tree; };
\end{lstlisting}
}
    \vspace{-0.2in}
\caption{The data structure of the range tree under our PAM interface. Due to space limit we only show the construction code.}
\label{fig:rangetree_code}
\end{figure}}

\hide{
  RangeQuery(vector<Point>& p) {
    const size_t n = p.size();
    pair<pnt, wght> *s = new pair<pnt, wght>[n];
    cilk_for (int i = 0; i < n; ++i)
      s[i] = make_pair(make_pair(p[i].x, p[i].y),
        p[i].w);
    range_tree.build(s, s + n, add);
    delete[] s;
  }	
}

To answer the query, we conduct two nested range searches: $(x_L, x_R)$ on the outer
tree, and $(y_L, y_R)$ on the related inner trees~\cite{de2000computational,samet1990design}. It can be implemented using the augmented map functions as:
{

\newdimen\zzsize
\zzsize=9pt
\newdimen\kwsize
\kwsize=9pt

\newcommand{\basicstyle}{\fontsize{\zzsize}{1.1\zzsize}\sc}
\newcommand{\keywordstyle}{\fontsize{\kwsize}{1.1\kwsize}\ttfamily\bf}

\newdimen\zzlstwidth
\settowidth{\zzlstwidth}{{\basicstyle~}}
\newcommand{\lcm}{}

\lstset{
  xleftmargin=5.0ex,
  basewidth=\zzlstwidth,
  basicstyle=\basicstyle,
  columns=fullflexible,
  captionpos=b,
  numbers=left, numberstyle=\small, numbersep=8pt,
  language=CAML,
  keywordstyle=\keywordstyle,
  keywords={signature,sig,structure,struct,fun,fn,case,type,datatype,and,let,fn,in,end,functor,alloc,if,then,else,while,with,and,start,do},
  commentstyle=\rmfamily\slshape,
  morecomment=[l]{\%},
  lineskip={1.5pt},
  columns=fullflexible,
  keepspaces=true,
  mathescape=true,
  escapeinside={@}{@},
  literate={requires}{{$\lcm\text{\keywordstyle \% requires}$}}6
           {returns}{{$\lcm\text{\keywordstyle \% returns}$}}6
           {=}{{$\lcm=$}}2 
           {(}{{$($}}2
           {)}{{$)$}}2 
           {**}{{$\lcm\times$}}2 
           {|}{{$|$}}2 
           {fn}{{$\lcm\boldsymbol\lambda\hspace{-1ex}$}}1 
           {==>}{{$\lcm\boldsymbol.\hspace{-1ex}$}}1 
           {->}{{$\lcm\rightarrow$}}2 
           {'a}{{$\alpha$}}1 
           {'b}{{$\beta$}}1 
}

\begin{lstlisting}[numbers=none]
query$(r_O,x_L,x_R,y_L,y_R)$ =
    let $g'(r_I)$ = AugRange$(r_I,y_L,y_R)$
    in AugProject$(g', +_{W}, r_O, x_L, x_R)$
\end{lstlisting}
  }
The \mb{augProject} on $R_O$ is the top-level searching of x-coordinates in the outer tree, and $g'$ projects the inner trees to the weight sum of the corresponding y-range. $f'$ (i.e., $+_W$) combines the weight of all results of $g'$ to give the sum of weights in the rectangle. When $f'$ is an addition, $g'$ returns the range sum, and $f$ is a \union{}, the condition $f'(g'(a),g'(b))=g'(a)+g'(b)=g'(a\cup b)=g'(f(a,b))$ holds, so \augProject{} is applicable.
Combining the two steps together, the
query time is $O(\log^2 n)$.    We can also answer range queries that
report all point inside a rectangle in time $O(k + \log^2 n)$, where $k$
is the output size.

\newcommand{\cfont}[1]{\textbf{#1}}
\newcommand{\cand}{\cfont{and}}
\newcommand{\cor}{\cfont{or}}
\newcommand{\candnot}{\cfont{and-not}}

\subsection{Ranked Queries on Inverted Indices}
\label{sec:index}

Our last application of augmented maps is building and searching
a weighted inverted index of the kind used by search
engines~\cite{ZM06,rajaraman2011} (also called an inverted file or
posting file).  For a given corpus, the index stores a mapping from
words to second-level mappings.  Each second-level mapping, maps each
document that the term appears in to a weight, corresponding to the
importance of the word in the document and the importance of the
document itself.  Using such a representation, conjunctions and
disjunctions on terms in the index can be found by taking the
intersection and union, respectively, of the corresponding maps.
Weights are combined when taking unions and intersections.  It is
often useful to only report the $k$ results with highest weight, as a search engine
would list on its first page.

This can be represented rather directly in our interface.  The inner
map, maps document-ids ($D$) to weights ($W$) and uses maximum as the
augmenting function $f$.  The outer map maps terms ($T$) to inner
maps, and has no augmentation.  This corresponds to the maps:

{\small
\begin{tabular}{l@{}@{ }l@{ }@{}r@{}@{ }l@{ }@{}l@{ }@{}l@{ }@{}l@{ }@{}l@{ }@{}l@{ }@{}l@{ }@{}l@{ }@{}l@{ }}
$M_I$ &$=$& $\mm$&$($&$D$,&$<_D$,&$W$,&$W$,&$(k,v) \mto v$, &${\max}_W$,&$0$&)\\
$M_O$ &$=$& $\mathbb{M}$&$($&$T$,&$<_T$&$M_I$,&$)$
\end{tabular}
\hide{\[M_I &=& \mm(D, W, <_D, W, (k,v) \mto v, {\max}_W, 0)\]
\[M_O &=& \mm(T, M_I, <_T ) .\]}
}

We use $\mathbb{M}(K,<_K,V)$ to represent a plain map with key type $K$, total ordering defined by $<_K$ and value type $V$. In the implementation, we use the feature of PAM that allows passing a combining function
with \union{} and \intersect{} (see Section \ref{sec:algorithm}), for combining weights.  The
\augFilter{} function can be used to select the $k$ best results
after taking unions and intersections over terms.  Note that an
important feature is that the \union{} function can take time
that is much less that the size of the output (e.g., see Section \ref{sec:algorithm}).  Therefore using
augmentation can significantly reduce the cost of finding the top $k$
relative to naively checking all the output to pick out the $k$ best.
The C++ code for our implementation is under 50 lines.

\section{Experiments}
\label{sec:exp}
For the experiments we use a 72-core Dell R930 with 4 x Intel(R)
Xeon(R) E7-8867 v4 (18 cores, 2.4GHz and 45MB L3 cache), and 1Tbyte memory.  Each core is
2-way hyperthreaded giving 144 hyperthreads.    Our code was compiled
using g++ 5.4.1, which supports the Cilk Plus extensions.
Of these we only use \ctext{cilk\_spawn} and \ctext{cilk\_sync} for
fork-join, and \ctext{cilk\_for} as a parallel loop.  We compile with
\ctext{-O2}. We use \ctext{numactl -i all} in all experiments with more than one thread.
It evenly spreads the memory pages across the processors in a round-robin fashion.

We ran experiments that measure performance of our four
applications: the augmented sum (or max), the interval tree, the 2D range tree and the word index
searching.

\newcolumntype{C}{>{$}c<{$}}
\newcolumntype{R}{>{$}r<{$}}
\setlength{\tabcolsep}{5pt}
\begin{table}[t]
\small
\begin{center}
\vspace{-0.1in}
\begin{tabular}{|@{ }l@{ }|@{ }C@{ }|@{ }C@{ }|R|R|R|}
\hline
            & \textbf{n} & \textbf{m} & \bf T_1 & \bf T_{144}&\textbf{Spd.} \\ \hline\hline
\multicolumn{6}{|@{ }l|}{\bf PAM (with augmentation)}\\ \hline
Union   & 10^8 & 10^8 & 12.517 & 0.2369 & 52.8\\ \hline
Union   & 10^8 & 10^5 & 0.257 & 0.0046 & 55.9 \\ \hline
Find              & 10^8 & 10^8 & 113.941 & 1.1923 & 95.6 \\ \hline
Insert            & 10^8 & - &205.970  & - &-\\ \hline
Build             & 10^8 & - & 16.089 & 0.3232 & 49.8\\ \hline
{\bf Build}             & {\bf 10^{10}} & - & {\bf 1844.38} & {\bf 28.24} & {\bf 65.3}\\ \hline
Filter            & 10^8 & - & 4.578 & 0.0804 &56.9\\ \hline
Multi-Insert      & 10^8 & 10^8 & 23.797 & 0.4528 & 52.6\\ \hline
Multi-Insert      & 10^8 & 10^5 & 0.407 &0.0071&57.3	 \\ \hline
Range      & 10^8 & 10^8 & 44.995 & 0.8033&	56.0\\ \hline
AugLeft &10^8 & 10^8& 106.096&	1.2133&	87.4\\ \hline
AugRange      & 10^8 & 10^8 & 193.229&	2.1966&	88.0\\ \hline
{\bf AugRange}      & {\bf 10^{10}} & {\bf 10^8} & {\bf 271.09}&	{\bf 3.04}	&{\bf 89.2}\\ \hline
AugFilter & 10^8 & 10^6 &0.807&0.0163&49.7\\
\hline
AugFilter & 10^8 & 10^5 &0.185 &0.0030 &61.2\\
\hline		
\multicolumn{6}{|@{ }l|}{\bf Non-augmented PAM (general map functions)}\\
\hline
Union   & 10^8 & 10^8 & 11.734 & 0.1967 & 59.7\\ \hline
Insert            & 10^8 & - & 186.649 & - &-\\ \hline
build             & 10^8 & - & 15.782 &0.3008	&52.5 \\ \hline
Range      & 10^8 & 10^8 &42.756 &0.7603 &	56.2\\ \hline\hline
\multicolumn{6}{|@{}l|}{\bf Non-augmented PAM (augmented functions)}\\
\hline
AugRange&10^8&10^4&21.642&0.4368&49.5\\
\hline
AugFilter&10^8&10^6&2.695 &0.0484&55.7\\
\hline
AugFilter&10^8&10^5&2.598&0.0497&52.3\\
\hline
\hline\hline
\multicolumn{6}{|@{ }l|}{\bf STL} \\ \hline
Union Tree    & 10^8 & 10^8 & 166.055 & - &-\\ \hline
Union Tree    & 10^8 & 10^5 & 82.514 & - &-\\ \hline
Union Array    & 10^8 & 10^8 & 1.033 & - &-\\ \hline
Union Array   & 10^8 & 10^5 & 0.459 & - &-\\ \hline
Insert        & 10^8 & - & 158.251 & - &-\\ \hline
\hline
\multicolumn{6}{|@{ }l|}{\bf MCSTL} \\ \hline
Multi-Insert    & 10^8 & 10^8 & 51.71 & 7.972 & 6.48\\ \hline
Multi-Insert    & 10^8 & 10^5 & 0.20 & 0.027 &7.36\\ \hline
\end{tabular}
\end{center}
\caption{Timings in seconds for various functions in PAM, the C++ Standard
  Template Library (STL) and the library Multi-core STL (MCSTL) \cite{SSP07}.  Here ``$T_{144}$'' means on all 72 cores with hyperthreads (i.e., 144 threads), and ``$T_1$'' means the same algorithm running on one thread. ``Spd.'' means the speedup (i.e., $T_1/T_{144}$). For insertion we test the total time of $n$ insertions in turn starting from an empty tree. All other libraries except PAM are not augmented.}\label{tab:genresult}
  \vspace{-.3in}
\end{table}

\begin{table}
  \centering\small
  \begin{tabular}{|@{}c@{}|@{}c@{}||@{}r|@{}r|@{}r||@{}r|@{}r|@{}r|}
  \hline
  & & \multicolumn{3}{@{}c@{}||}{\textbf{Overhead for aug.}} &\multicolumn{3}{c|}{\textbf{Saving from node-sharing}}\\
  \cline{3-8}
  \textbf{Func.}&\textbf{Type}& \textbf{node}  & \textbf{aug.} & \textbf{over-}&\textbf{\#nodes in} &\textbf{Actual}& \textbf{Saving}\\
  &&\textbf{size}&\textbf{size}&\textbf{head} & \textbf{theory}& \textbf{\#nodes}&\textbf{ratio}\\
    \hline
    \multirow{2}{*}{\textbf{Union}} & $\bf m=10^8$ & 48B & 8B & $20\%$ & $390$M& $386$M & $1.2\%$ \\
    \cline{2-8}
     & $\bf m=10^5$& 48B & 8B & $20\%$ & $200$M& $102$M& $49.0\%$ \\
    \hline
{\textbf{Range}} & \textbf{Outer} & 48B & 8B & $20\%$ & $100$M& $100$M& $0.0\%$ \\
    \cline{2-8}
   \textbf{Tree}  & \textbf{Inner}& 40B & 4B & $11\%$ & $266$M& $229$M& $13.8\%$ \\
    \hline
  \end{tabular}
  \caption{Space used by the \union{} function and the range tree application. We use B for byte, M for million. }\label{tab:space}\vspace{-1em}
\end{table}
\hide{The overhead for augmentation is computed by aug\_size/(node\_size-aug\_size). The saving ratio from persistence is computed by 1 - \#actual/\#in\_theory.}

\begin{figure*}[!h!t]
\centering
\begin{tabular}{ccc}
\multicolumn{3}{c}{\begin{tabular}{cc}
{(a). $5\times 10^7$ insertions, throughput (M/s), $p=144$.}&(b). $10^7$ concurrent reads, throughput (M/s), $p=144$.\\
Compare to some concurrent data structures.&Compare to some concurrent data structures.\\
\includegraphics[width=0.32\columnwidth]{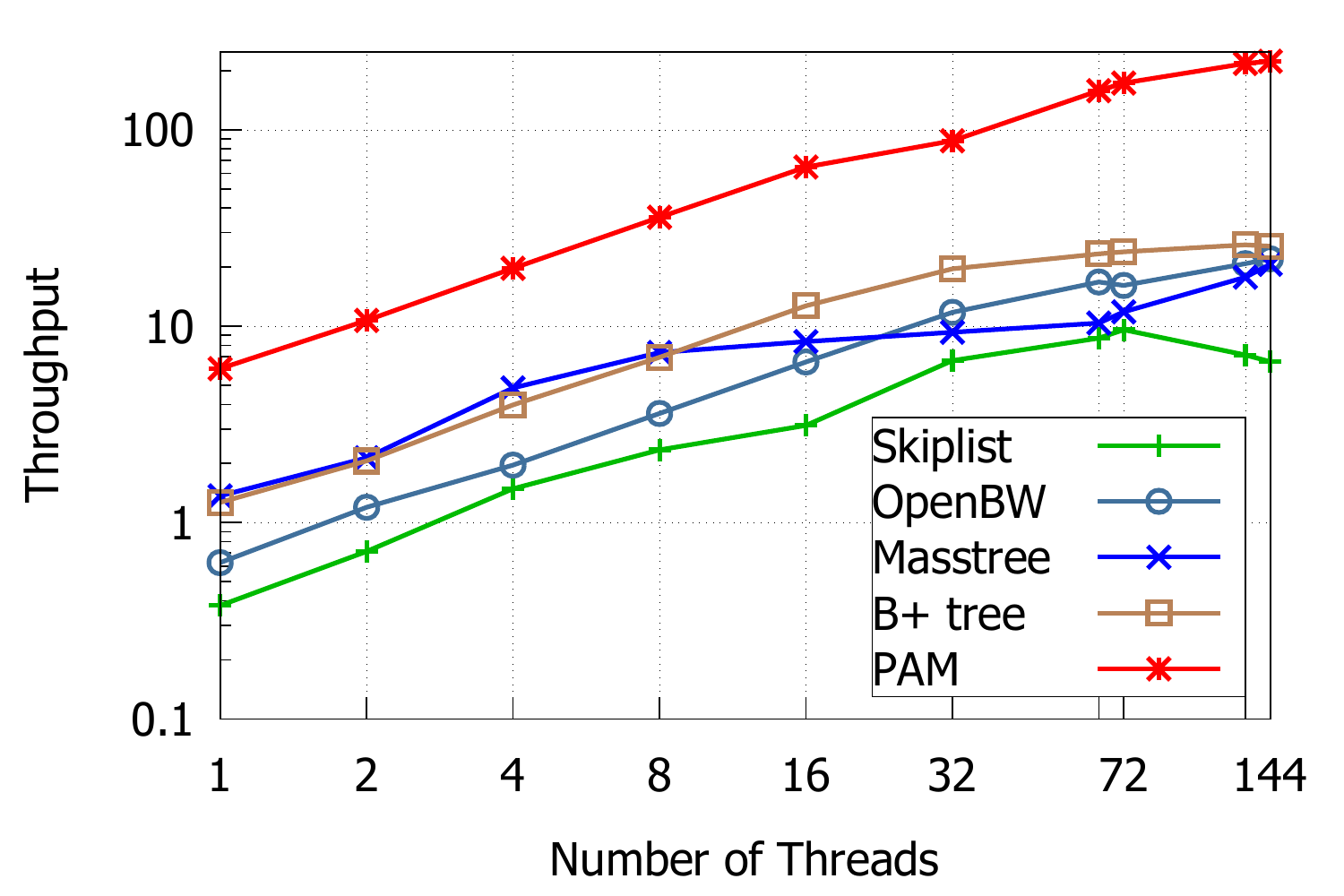}&
\includegraphics[width=0.32\columnwidth]{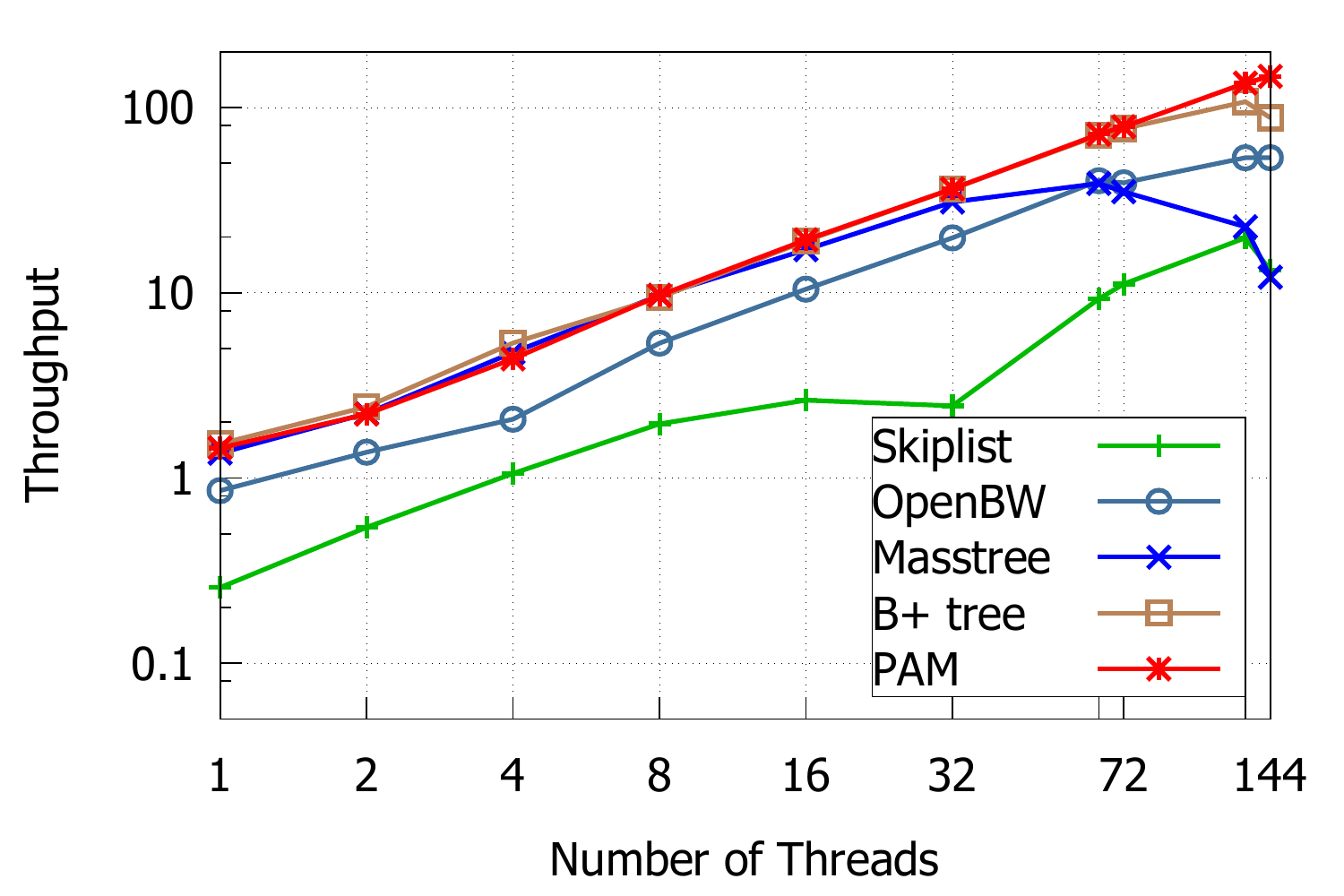}\\
\end{tabular}}\\
(c). $n=10^8$, running time.&(d). $n=10^8$, speedup.&(e). $n=10^8$, sequential building time.\\
PAM on different input sizes. &The interval tree using PAM.&The range tree. Compare to CGAL.\\
\includegraphics[width=0.32\columnwidth]{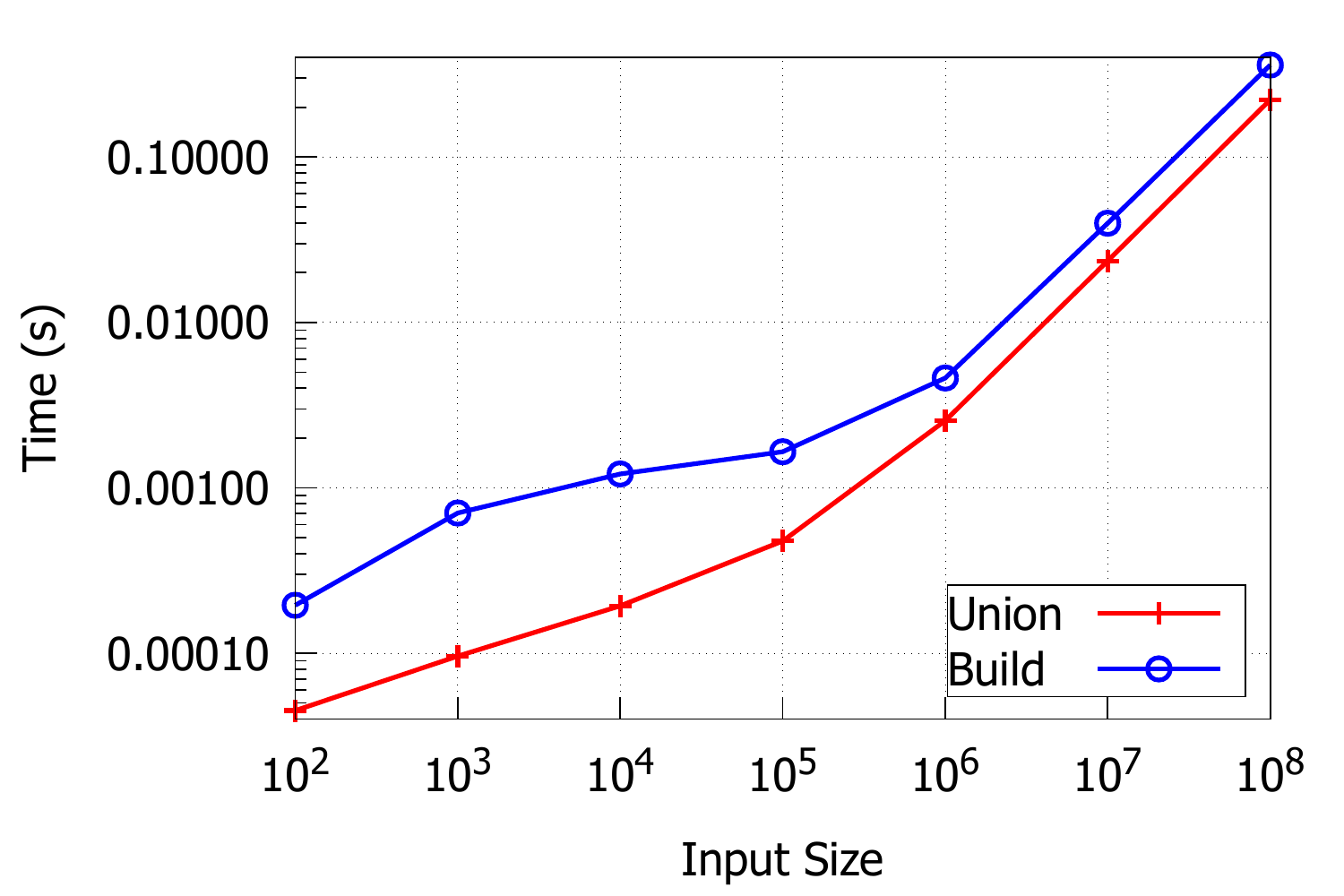}&
\includegraphics[width=0.32\columnwidth]{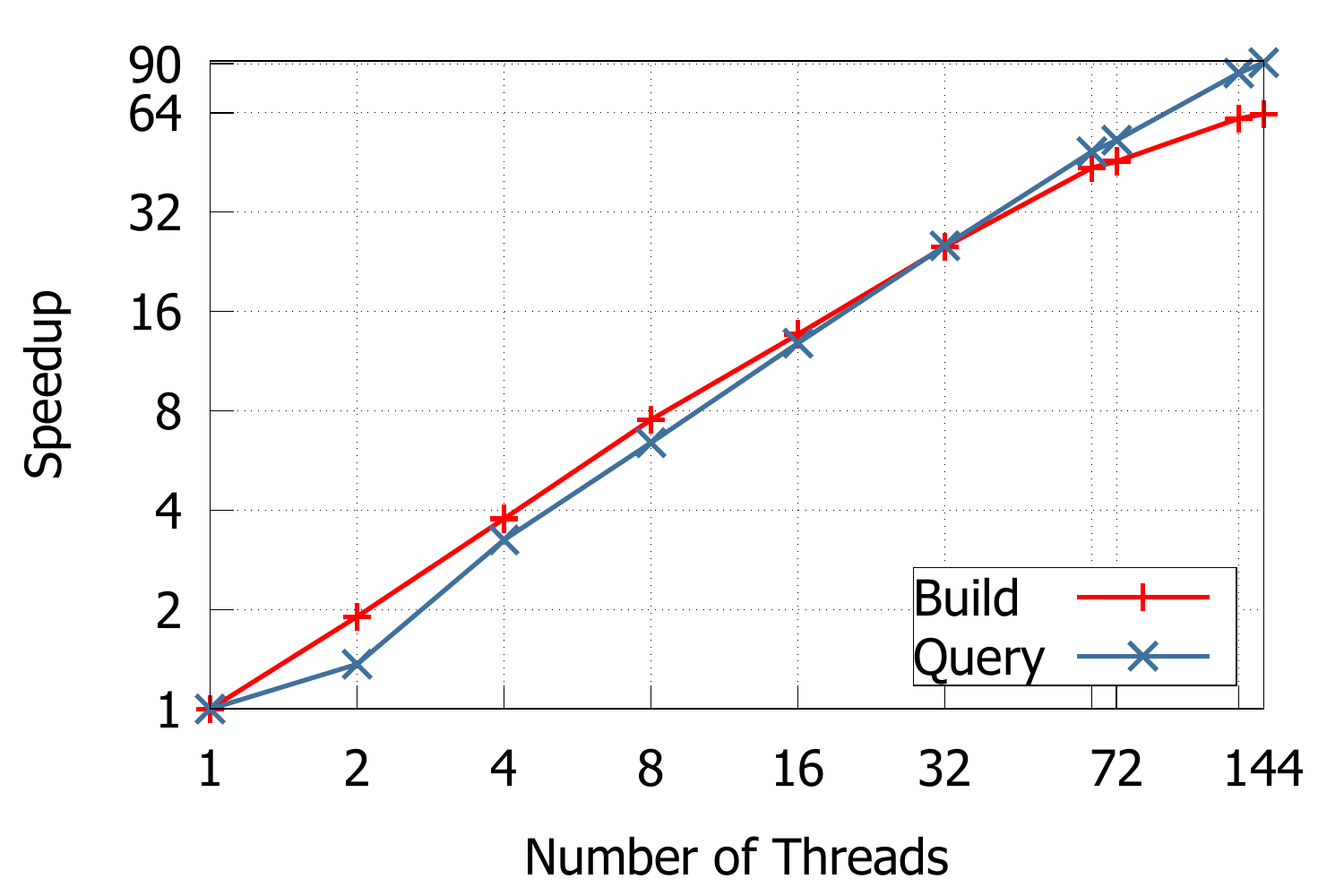}&
\includegraphics[width=0.32\columnwidth]{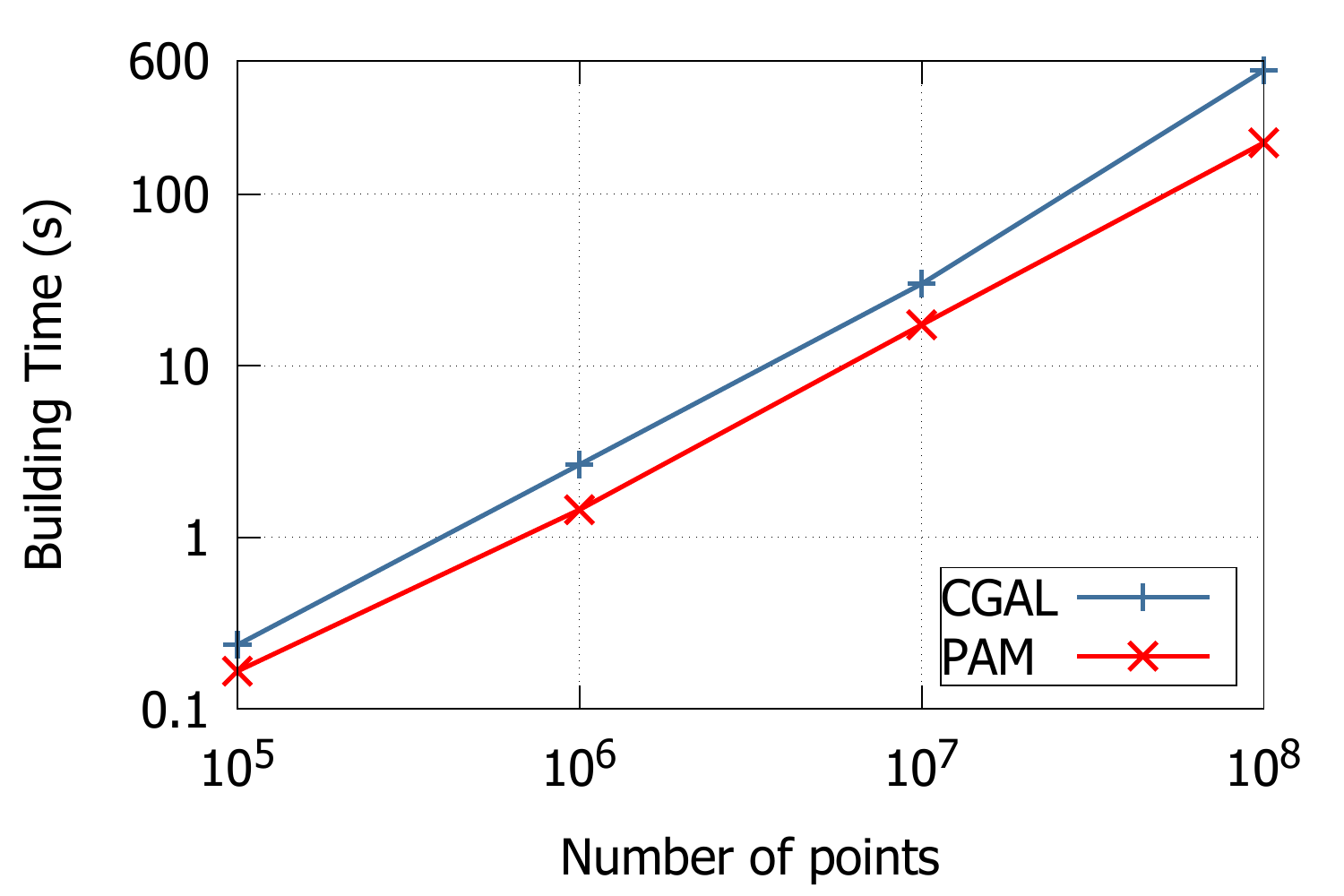}\\
\end{tabular}
\vspace{-0.1in}
  \caption{(a), (b) The performance (throughput, millions of elements per second) of PAM comparing with some concurrent data structures. In (a) we use our \multiInsert{}, which is not as general as the concurrent insertions in other implementations. (c) The running time of \union{} and \build{} using PAM on different input sizes. (d) The speedup on interval tree construction and query. (e) The running time on range tree construction.}\label{fig:exps}
  \vspace{-0.1in}
\end{figure*}
\subsection{The Augmented Sum}
Times for set functions such as union using the same algorithm in PAM
without augmentation have been summarized in~\cite{ours}. In this
section we summarize times for a simple augmentation, which just adds
values as the augmented value (see Equation \ref{eqn:exampleeqn}).  We
test the performance of multiple functions on this structure.  We also
compare PAM with some sequential and parallel libraries, as well as
some concurrent data structures. None of the other implementations
support augmentation.  We use 64-bit integer keys and values.  The
results on running time are summarized in Table~\ref{tab:genresult}.
Our times include the cost of any necessary garbage collection
(GC). We also present space usage in Table \ref{tab:space}.

We test versions both with and
without augmentation. For general map functions like \union{} or \insertnew{},
maintaining the augmented value in each node costs overhead, but it
seems to be minimal in running time (within 10\%).  This is likely
because the time is dominated by the number of cache misses, which is
hardly affected by maintaining the augmented value. The overhead of space in maintaining the augmented value is 20\% in each tree node (extra 8 bytes for the augmented value).
For the functions related to the range sum,
the augmentation is necessary for theoretical efficiency, and greatly improves the performance. For example, the \augRange{} function
using a plain (non-augmented) tree structure would require scanning all entries in the range, so the
running time is proportional to the number of related entries. It costs 0.44s to process $10^4$ parallel \augRange{} queries.
With augmentation, \augRange{} has
performance that is close to a simple
\find{} function, which is only 3.04s for $10^8$ queries. Another example to show the advantage of augmentation is the \augFilter{} function. Here we use \func{Max} instead of taking the sum as the combine function, and set the filter function as selecting all entries with values that are larger than some threshold $\theta$. 
We set the parameter $m$ as the output size, which can be adjusted by choosing appropriate $\theta$. Such an algorithm has theoretical work of $O(m\log (n/m+1))$, and is significantly more efficient than a plain implementation (linear work) when $m\ll n$. We give two examples of tests on $m=10^5$ and $10^6$. The change of output size does not affect the running time of the non-augmented version, which is about 2.6s sequentially and 0.05s in parallel. When making use of the augmentation, we get a 3x improvement when $m=10^6$ and about 14x improvement when $m=10^5$.

For sequential performance we compare to the C++ Standard Template
Library (STL)~\cite{musser2009stl}, which supports
\texttt{set\_union} on sets based
on red-black trees and sorted vectors (arrays). We denote the two versions as
\emph{Union-Tree} and \emph{Union-Array}.
In Union-Tree results are inserted into a new tree, so it is also persistent.
When the two sets have the same size, the
array implementation is faster because of its flat structure and
better cache performance. If one map is much smaller, PAM performs better than Union-Array
because of better theoretical bound ($O(m\log (n/m+1))$ vs. $O(n+m)$). It outperforms Union-Tree because it supports persistence
more efficiently, i.e., sharing nodes instead of making a copy of all output entries.
Also, our \join{}-based \insertnew{} achieves performance close to (about $17\%$ slower) the
well-optimized STL tree insertion even though PAM needs to maintain the
reference counts.

In parallel, the speedup on the aggregate functions such as \union{}
and \build{} is above 50.
Generally, the speedup is correlated to the ratio of reads to writes.
With all (or mostly) reads to the tree structure, the speedup is often more than 72 (number of cores) with hyperthreads (e.g., \find{}, \augLeft{} and \augRange{}).
With mostly writes (e.g., building a new tree as output) it is 40-50 (e.g., \filter{}, \range{}, \union{} and \augFilter{}). The \build{} function is relatively special because the parallelism is mainly from the parallel sorting.
We also give the performance of the
\multiInsert{} function in the Multicore STL (MCSTL) \cite{SSP07}
for reference.  On our server MCSTL does not scale to 144 threads, and
we show the best time it has (on 8-16 threads). On the functions we
test, PAM outperforms MCSTL both sequentially and in parallel.

PAM is scalable to very large data, and still achieve very good speedup. On our machine, PAM can process up to $10^{10}$ elements (highlighted in Table \ref{tab:genresult}). 
It takes more than half an hour to build the tree sequentially, but only needs 28 seconds in parallel, achieving a 65-fold speedup. For \augRange{} the speedup is about 90.

Also, using path-copying to implement persistence improves
space-efficiency.  For the persistent \union{} on two maps of size
$10^8$ and $10^5$, we save about $49\%$ of tree nodes because most nodes in
the larger tree are re-used in the output tree.  When the two trees are of the same size
and the keys of both trees are extracted from the similar distribution, there is little savings.


We present the parallel running times of \union{} and \build{} on
different input sizes in Figure \ref{fig:exps} (c). For \union{} we
set one tree of size $10^8$ and vary the other tree size. When the
tree size is small, the parallel running time does not shrink
proportional to size (especially for \build{}), but is still
reasonably small.  This seems to be caused by insufficient parallelism
on small sizes.  When the input size is larger than $10^6$, the
algorithms scales very well.

We also compare with four comparison-based concurrent data structures: skiplist, OpenBw-tree \cite{bwtree}, Masstree \cite{mao2012cache} and B+ tree \cite{zhang2016reducing}. The implementations are from \cite{bwtree}\footnote{We do not compare to  the fastest implementation (the Adaptive Radix Tree \cite{leis2013adaptive}) in \cite{bwtree} because it is not comparison-based.}. We compare their concurrent insertions with our parallel \multiInsert{} and test on YCSB microbenchmark C (read-only). We first use $5\times10^7$ insertions to an empty tree to build the initial database, and then test $10^7$ concurrent reads. The results are given in Figure \ref{fig:exps}(a) (insertions) and (b) (reads). For insertions, PAM largely outperforms all of them sequentially and in parallel, although we note that their concurrent insertions are more general than our parallel multi-insert (e.g., they can deal with ongoing deletions at the same time). For concurrent reads, PAM performs similarly to B+ tree and Masstree with less than 72 cores, but outperforms all of them on all 144 threads. We also compare to Intel TBB~\cite{pheatt2008intel,TBB} concurrent hash map, which is a parallel implementation on \emph{unordered} maps. On inserting $n=10^8$ entries into a pre-allocated table of appropriate size, it takes 0.883s compared to our 0.323s (using all 144 threads).

\subsection{Interval Trees}
We test our interval tree (same code as in Figure \ref{fig:interval_tree}) using the PAM library. For queries we test $10^9$ stabbing queries. We give the results of our interval tree on $10^8$ intervals in Table \ref{tab:rt-exp} and
the speedup figure in Figure \ref{fig:exps}(d).

\hide{
We compare the sequential performance of our interval tree with a Python interval tree implementation~\cite{IntervalPython}. The Python implementation is sequential, and is very inefficient. We ran the Python interval tree only up to a tree size of $10^7$ because it already took about $200$s to build the tree when the size is $10^7$, and can hardly accept larger trees as input. The sequential running time of our code and the Python interval tree is reported in Table \ref{tab:interval}. In building the tree, our code is about dozens of times more efficient than the Python implementation, and in performing queries, is orders of magnitude faster. Our code can answer millions of queries in one second, while the Python interval tree can only do $418$ per second when $n$ is as small as $10^4$, and this processing rate deceases dramatically when $n$ gets larger. The sequential running time of our implementation is also given in Figure \ref{fig:it-exp}(a).}

Sequentially, even on $10^8$ intervals the tree construction only takes 14 seconds, and each query takes around 0.58 $\mu$s.  We did not find any comparable open-source interval-tree library in C++ to compare with. The only available library is a Python interval tree implementation~\cite{IntervalPython}, which is sequential, and is very inefficient (about 1000 times slower sequentially).
Although unfair
to compare performance of C++ to python (python is optimized for ease
of programming and not performance), our interval tree is much simpler than the python code---30 lines in
Figure~\ref{fig:interval_tree} vs. over 2000 lines of python.  This
does not include our code in PAM (about 4000
lines of code), but the point is that our library can be shared among many
applications while the Python library is specific for the interval query.  Also our code supports parallelism.

In parallel, on $10^8$ intervals, our code can build an interval tree
in about 0.23 second, achieving a 63-fold speedup. We also give the
speedup of our PAM interval tree in Figure \ref{fig:exps}(d). Both
construction and queries scale up to 144 threads (72 cores with
hyperthreads).

\subsection{Range Trees}
We test our 2D range tree as described in Section \ref{sec:rangetreeapp}.
A summary of run times is presented in Table \ref{tab:rt-exp}.
We compared our sequential version with the range tree in the CGAL
library~\cite{overmars1996designing} (see Figure \ref{fig:exps}(e)).
The CGAL range tree is sequential and can only report all the points in the
range.   For $10^8$ input points we control
the output size of the query to be around $10^6$ on average.
Table~\ref{tab:rt-exp} gives results of construction and query time using PAM and CGAL respectively. Note that our range tree also stores the weight and reference counting in the tree while CGAL only stores the coordinates, which means CGAL version is more space-efficient than ours. Even considering this, PAM is always more efficient than CGAL and less than half the running time both in building and querying time on $10^8$ points. Also our code can answer the weight-sum in the window in a much shorter time, while CGAL can only give the full list.

We then look at the parallel performance. As shown in Table
\ref{tab:rt-exp} it took $3$ seconds (about a 64-fold speedup)
to build a tree on $10^8$ points. On $144$ threads the PAM range tree can
process $1.82$ million queries on weight-sum per second, achieving a
87-fold speedup.

We also report the number of allocated tree nodes in Table \ref{tab:space}. Because of path-copying, we
save 13.8\% space by the sharing of inner tree nodes.


\begin{table}[t]
\small
  \centering\small
\begin{tabular}{|l|l||c|c||r|r||c|}
\hline
   {\textbf{Lib.}}& {\textbf{Func.}}    &{\textbf{n}}&{\textbf{m}}& $\bf T_1$ & $\bf T_{144}$ & \textbf{Spd.}\\
   \hline
\multirow{1}{*}{\textbf{PAM}}&\textbf{Build}& $10^8$& -  & 14.35  &0.227	&	63.2 \\ \cline{2-7}
\multirow{1}{*}{\textbf{(interval)}}&\textbf{Query}& $10^8$ & $10^8$ & 53.35 &0.576& 92.7\\ \hline
\multirow{2}{*}{\textbf{PAM}}&\textbf{Build}& $10^8$& -  &   197.47&	3.098&	63.7 \\ \cline{2-7}
\multirow{2}{*}{\textbf{(range)}}&\textbf{Q-Sum}& $10^8$ & $10^6$ & 48.13&	0.550	&87.5 \\ \cline{2-7}
&\textbf{Q-All}& $10^8$ & $10^3$ &  44.40  &   0.687 & 64.6 \\ \hline
\multirow{1}{*}{\textbf{CGAL}}&\textbf{Build} & $10^8$ & - & 525.94 &- & -\\ \cline{2-7}
\multirow{1}{*}{\textbf{(range)}}&\textbf{Q-All} & $10^8$ & $10^3$ & 110.94 &- & - \\ \hline
\end{tabular}
\vspace{.06in}
  \caption{The running time (seconds) of the range tree and the interval tree implemented with PAM interface on $n$ points and $m$ queries.  Here ``$T_1$'' reports the sequential running time and ``$T_{144}$'' means on all 72 cores with hyperthreads (i.e., 144 threads). ``Spd.'' means the speedup (i.e., $T_1/T_{144}$). ``Q-Sum'' and ``Q-All'' represent querying the sum of weights and querying the full list of all points in a certain range respectively. We give the result on CGAL range tree for comparisons with our range tree.}\label{tab:rt-exp}
  \vspace{-0.2in}
\end{table}

\subsection{Word Index Searching}

To test the performance of the inverted index data structure described
in Section~\ref{sec:index}, we use the publicly available Wikipedia
database~\cite{Wikipedia} (dumped on Oct.\ 1, 2016) consisting of 8.13
million documents.  We removed all XML markup, treated everything
other than alphanumeric characters as separators, and converted all
upper case to lower case to make searches case-insensitive.  This
leaves 1.96 billion total words with 5.09 million unique words.  We
assigned a random weight to each word in each document (the values of
the weights make no difference to the runtime).  We measure the
performance of building the index from an array of \texttt{(word,
  doc\_id, weight)} triples, and the performance of queries that take
an intersection (logical-and) followed by selecting the top 10 documents
by weight.

Unfortunately we could not find a publicly available C++ version of
inverted indices to compare to that support and/or queries and
weights although there exist benchmarks supporting plain searching on a single word~\cite{invertedindex}.
However the experiments do demonstrate speedup numbers,
which are interesting in this application since it is the only one
which does concurrent updates.  In particular each query does its own
intersection over the shared posting lists to create new lists (e.g.,
multiple users are searching at the same time).
Timings are shown in Table~\ref{fig:word}.  Our implementation can
build the index for Wikipedia in 13 seconds, and can answer 100K
queries with a total of close to 200 billion documents across the
queries in under 5 seconds.  It demonstrates that good speedup (77x)
can be achieved for the concurrent updates in the query.

\begin{table}[t]
\begin{center}
\small
\begin{tabular}{|@{ }l@{ }||c||c|c||c|c||c|}
\hline
       & \multirow{2}{*}{\textbf{n}} & \multicolumn{2}{c||}{\textbf{1 Core}} & \multicolumn{2}{c||}{\textbf{$\bf{72^*}$ Cores}} & \multirow{2}{*}{\textbf{Speed-}} \\ \cline{3-6}
       &  \multirow{2}{*}{($\bf \times 10^9$)}   & \textbf{Time} & \textbf{Melts} & \textbf{Time}& \textbf{Gelts} &\multirow{2}{*}{\textbf{up}}\\
       &    & \textbf{(secs)} & \textbf{/sec} & \textbf{(secs)} & \textbf{/sec}& \\ \hline
\textbf{Build}   &  1.96 &  1038  & 1.89  &  12.6 & 0.156  &82.3 \\ \hline
\textbf{Queries}  & 177 &  368  &  480.98  &  4.74 & 37.34  & 77.6\\ \hline
\end{tabular}
\end{center}
\caption{The running time and rates for building
and queering an inverted index. Here ``one core'' reports the sequential running time and ``$\bf 72^*$ cores'' means on all 72 cores with hyperthreads (i.e., 144 threads). Gelts/sec calculated as $n/(\mbox{time} \times 10^9)$.}
\label{fig:word}
\vspace{-.3in}
\end{table}

\section{Conclusion}
\label{sec:conclusion}
In this paper we introduce the \emph{augmented map}, and describe an
interface and efficient algorithms to for it.  Based on the interface
and algorithms we develop a library supporting the augmented map
interface called PAM, which is parallel, work-efficient, and supports
persistence.  We also give four example applications that can be
adapted to the abstraction of augmented maps, including the augmented
sum, interval trees, 2D range trees and the inverted indices. We
implemented all these applications with the PAM library. Experiments
show that the functions in our PAM implementation are efficient both
sequentially and in parallel. The code of the applications implemented
with PAM outperforms some existing libraries and implementations, and
also achieves good parallelism. Without any specific optimizations,
the speedup is about more than 60 for both building interval trees and
building range trees, and 82 for building word index trees on 72
cores. For parallel queries the speedup is always over 70.

\bibliographystyle{plainnat}
\bibliography{main}

\end{document}